\documentclass[12pt]{article}
\usepackage{amsmath,amsfonts}
\usepackage{hyperref}
\usepackage{graphicx}
\usepackage{slashed}

\usepackage[numbers,sort&compress]{natbib}

\usepackage{amsthm}
\usepackage{amssymb}
\usepackage[matrix,arrow]{xy}
\usepackage{epsfig}
\usepackage{ytableau}
\ytableausetup{boxsize=.6em,centertableaux}
\makeatletter\@addtoreset{equation}{section}\makeatother

\def\Z {\mathbb{Z}}
\newcommand{\anti}{\ydiagram{1,1}}
\newcommand{\sym}{\ydiagram{2}}

\newcommand{\al}{\alpha}
\newcommand{\be}{\beta}
\newcommand{\te}{\theta}
\newcommand{\la}{\lambda}

\newcommand{\onov}[1]{\frac{1}{#1}}
\newcommand{\mat}[1]{\left(\begin{matrix} #1 \end{matrix}\right)}

\newcommand{\lag}{\mathcal{L}}

\newcommand{\beq}{\begin{equation}}
	\newcommand{\eeq}{\end{equation}}
\newcommand{\bea}{\begin{eqnarray}}
	\newcommand{\eea}{\end{eqnarray}}

\newcommand{\vev}[1]{{\left< {#1} \right>}}

\newcommand{\eql}[2]{\begin{equation}\label{#1}{\begin{split}#2\end{split}}\end{equation}}
\newcommand{\eq}[2][ ]{\begin{equation}\label{#1}{\begin{split}#2\end{split}}\end{equation}}

\newcommand{\tr}{{\rm tr\,}}

\renewcommand{\title}[1]{\vbox{\center\LARGE{#1}}\vspace{5mm}}
\renewcommand{\author}[1]{\vbox{\center\large#1}\vspace{5mm}}
\newcommand{\address}[1]{\vbox{\center\em#1}}
\newcommand{\email}[1]{\vbox{\center\tt#1}\vspace{5mm}}

\begin{document}

\title{Anomalies for anomalous symmetries}

\author{Avner Karasik}

\address{Department of Applied Mathematics and Theoretical Physics \\ University of Cambridge\\ CB3 0WA, UK}

\email{avnerkar@gmail.com}

\abstract{4d gauge theories with massless fermions typically have axial U(1) transformations that suffer from the ABJ anomaly. One can modify the theory of interest by adding more fields in a way that restores the axial symmetry, and use it to derive rigorous 't-Hooft anomaly matching conditions. These conditions are not valid for the original theory of interest, but for the modified theory. We show that the modification can be done in a specific way that allows us to relate the dynamics of the modified theory to the dynamics of the original theory. In this way, the anomaly matching conditions of the modified theory can be used to learn new things on the original theory even though they involve axial transformations which are not a symmetry of the original theory. We describe this method and discuss some applications to various examples.}

	\bibliographystyle{utphys}

\newpage
	\tableofcontents
\newpage
\section{Introduction}
Anomaly matching conditions, originally introduced by 't-Hooft \cite{tHooft:1979rat} is one of the most important theoretical tools in analysing strongly interacting quantum field theories. The idea can be described as follows. Consider a theory $\mathcal{T}_{uv}$ with some global symmetry group $G$ defined at high energies. We want to classify the set of possible low energy effective theories, $\mathcal{T}_{uv}$ can flow to. One thing that can be done is to couple the symmetry group $G$ to background gauge fields $A$, such that the partition function $\mathcal{Z}[A]$ depends on $A$. It happens to be that even though $G$ is a perfectly good symmetry of $\mathcal{T}_{uv}$, the partition function may not be completely gauge invariant. Instead, under gauge transformations $A\to A'$, the partition function may transform as
\eql{anomaly}{\mathcal{Z}[A]\to\mathcal{Z}[A']=e^{iS[A]}\mathcal{Z}[A]\ ,}
where $S[A]$ is a local action of the background gauge fields.\footnote{Of course, the phase of the partition function can be modified by adding local counter terms. As always when talking about anomalies, $S[A]$ is a phase that cannot be removed and hence corresponds to an anomaly.} The importance of $S[A]$ is that it is RG invariant. \eqref{anomaly} must be satisfied by the partition function at any scale. In particular, the low energy effective theory must reproduce the same phase. This gives a non-trivial constraint on the set of allowed low energy theories $\{\mathcal{T}_{IR}\}$. The basic ingredient in this construction is the symmetry group $G$. Without symmetries there will be no anomalies. Needless to say, fields transformations which are not symmetries of the theory, cannot be coupled to background gauge fields and cannot be used to derive rigorous constraints on the flow of the theory. 
In this work we will argue that the last statement is not entirely true. In some cases, rigorous anomaly-like constraints can be derived for certain discrete $\Z_N$ transformations which are not a symmetry of the theory. The constraints are much weaker than usual anomaly constraints, but they still teach us something non-trivial about strongly interacting QFTs.

\begin{figure}
	\vspace{4pt}
	\begin{center}
		\includegraphics[width=1\textwidth]{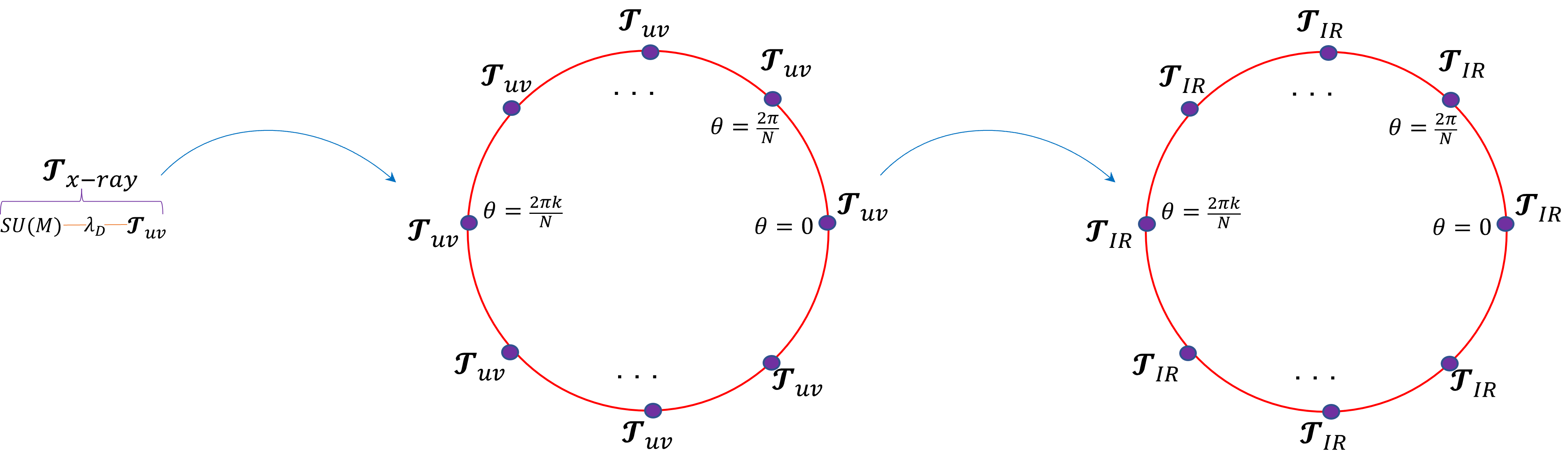}
		\caption{\footnotesize{This figure represents the flow of the theory. We start at high energies with $\mathcal{T}_{x-ray}$ which is a modification of $\mathcal{T}_{uv}$ by an extra gauge group $SU(M)$ and a fermion $\la_D$. Confinement of $SU(M)$ leads to a spontaneous breaking of the axial $\Z_N$ symmetry. In each vacuum we have $\mathcal{T}_{uv}$ with a different $\te$ angle. In the presence of massless fermions, the $\te$ angle can be rotated away, with the price of generating different background $\te$ terms for global symmetries in each vacuum. At low energies, $\mathcal{T}_{uv}$ flows to $\mathcal{T}_{IR}$. The demand that $N$ copies of $\mathcal{T}_{IR}$ will be consistent with the anomalies of $\mathcal{T}_{x-ray}$ put constraints on the flow of $\mathcal{T}_{uv}$. }}\label{flow}
	\end{center}
	\vspace{4pt}
\end{figure}

For concreteness, we will restrict attention to 4d gauge theories. In the presence of charged massless fermions, the uv Lagrangian of these theories is typically invariant under axial $U(1)_A$ transformations that act differently on left and right handed fermions. However, these transformations change the path integral measure and hence don't correspond to a symmetry of the full quantum theory. This is known as the Adler–Bell–Jackiw (ABJ) anomaly\cite{Adler:1969gk,Bell:1969ts}\footnote{In order to avoid confusions, from now on we will use the word "anomaly" only in the context of 't-Hooft anomalies.}. As a result, these $U(1)_A$ transformations don't contribute to anomaly matching conditions. 
Instead of studying directly the theory of interest $\mathcal{T}_{uv}$ (for which $U(1)_A$ is not a symmetry), we suggest to embed $\mathcal{T}_{uv}$ inside a bigger theory defined at higher energies, $\mathcal{T}_{x-ray}$. $\mathcal{T}_{x-ray}$ can be chosen such that it satisfies the following properties:
\begin{enumerate}
	\item There exists a discrete subgroup $\Z_N\subset U(1)_A$ which is an exact symmetry of $\mathcal{T}_{x-ray}$. 
	\item $\Z_N$ carries some anomalies and gives rigorous constraints on the flow of $\mathcal{T}_{x-ray}$.
	\item At intermediate energies, $\Z_N$ is spontaneously broken. The effective theory in each one of the $N$ vacua is exactly $\mathcal{T}_{uv}$. The anomalies associated with $\Z_N$ are carried partially by the condensate that breaks the symmetry, and partially by the field content of $\mathcal{T}_{uv}$.
	\item When sitting in one of the $N$ vacua, the flow of $\mathcal{T}_{x-ray}$ is completely equivalent to the flow of $\mathcal{T}_{uv}$. 
	\item The anomaly constraints associated with $\Z_N$ must be satisfied by the low energy theory of $\mathcal{T}_{x-ray}$. Because their flow is equivalent, the same conditions constrain the low energy theory of $\mathcal{T}_{uv}$. This is true even though $\Z_N$ is \textbf{not} a symmetry of $\mathcal{T}_{uv}$.
	\item Another way to view the last point: The dynamics that control the flow of $\mathcal{T}_{uv}$ don't know if the full theory is simply $\mathcal{T}_{uv}$ or whether it has some "x-ray completion" into a bigger theory. Hence, the dynamics must be consistent with all the possible embeddings of $\mathcal{T}_{uv}$, and in particular with $\mathcal{T}_{x-ray}$. 
\end{enumerate}
 It is important to emphasize that in the "uv" where $\mathcal{T}_{uv}$ is defined, the $\Z_N$ symmetry is already broken by a condensate made out of "x-ray" fields. However, if this condensate doesn't carry all the anomalies, then some are left to be matched by $\mathcal{T}_{uv}$ and therefore also by $\mathcal{T}_{IR}$. 
 Since the symmetry is broken already in the uv, the conditions we get are much weaker than those coming from usual anomalies. They cannot rule out phases, but they can still teach us non-trivial things on $\mathcal{T}_{IR}$.
  There are three types of anomalies that involve $\Z_N$ in 4d:
 \begin{enumerate}
 	\item $\Z_N\times G^2$: The first is a triangle anomaly linear in the broken $\Z_N$. Here $G$ stands for any symmetry transformation of $\mathcal{T}_{uv}$. In a phase where $G$ is unbroken, the anomaly is typically carried by a condensate charged under $\Z_N$. One can use the anomaly to argue that the phase of the condensate must couple to background gauge fields for $G$ in a way that reproduces the anomaly. There are two options,
 	\begin{itemize}
 		\item  If $G^2$ corresponds to a 3d anomaly, the anomaly will be carried by the effective 3d theory on domain walls connecting two distinct vacua. In other words, the dynamical 3d domain wall theory has the $G^2$ anomaly. An example for that is when $G$ is some $\Z_K$ one-form symmetry. We will see an example for this in section \ref{subsec:cusp}.
 		\item The second option is that $G^2$ corresponds not to a 3d anomaly but to a 2d anomaly. Then by definition, the domain wall theory cannot have the $G^2$ anomaly. One can construct a junction of domain walls. The 2d theory living on the junction then carries the $G^2$ anomaly. If we think of the $\Z_N$ as embedded inside a $U(1)$, the junction configuration is the same as a vortex-string of the $U(1)$ case, and the anomaly is matched by its effective 2d theory as in \cite{Callan:1984sa}. Typical examples for this is when $G^2$ is gravity, $U(1)^2$ or $SU(N_f)^2$. 
 	\end{itemize}
 	
 	\item $\Z_N^2\times G$: Anomalies quadratic in the discrete broken symmetry are also realized on junctions. The 2d theory on the junction should have a $\Z_N\times G$ anomaly. This anomaly cannot be realized on a domain wall simply because the $\Z_N$ is not a symmetry of the domain wall. On the junction, the $\Z_N$ is restored. We will see an example for this in section \ref{subsec:Nf1baryons}.
 	\item $\Z_N^3$: Exactly as in the previous case, triangle $\Z_N$ anomalies are realized as a $\Z_N^2$ anomaly on the junction. See appendix \ref{sec:junctionanomalies} for more details on anomaly matching on junctions.
 \end{enumerate} 
As we see, given that the $\Z_N$ symmetry is broken, the main thing these anomalies teach us regards constraining effective theories on solitons such as domain walls and junctions. In our setup, the $\Z_N$ anomalies constrain the effective theory on solitons of $\mathcal{T}_{x-ray}$. As we will argue, thanks to the scale separation and the dynamical relation between $\mathcal{T}_{x-ray}$ and $\mathcal{T}_{uv}$, the same anomalies can constrain the effective theory on solitons also for $\mathcal{T}_{uv}$ even though $\Z_N$ is not a symmetry of $\mathcal{T}_{uv}$. 
Before going into the details, we will give another point of view for why this procedure works. Axial rotations are transformations that change the path integral measure in a way which is equivalent to shifting the $\te$ term. While continuous axial rotations don't generate a symmetry, there might be discrete axial rotations that change $\te$ by $2\pi\Z$ and therefore leave the theory invariant. These discrete transformations are exact symmetries of the theory and can be used to derive anomaly constraints. The new idea presented in this work is to construct a setup in which $\te$ is effectively not $2\pi$ periodic but $2\pi/N$ periodic. This will immediately increase the axial symmetry of the theory by a factor of $\Z_N$. 
As will be shown in detail in section \ref{sec:YM}, this can be achieved by adding to the theory of interest $\mathcal{T}_{uv}$, another gauge group and a massless Dirac field in the bifundamental representation. This bifundamental theory is what we call $\mathcal{T}_{x-ray}$. In the limit where the new gauge group confines when the original gauge group is still weakly coupled, the effective theory after the confinement of the new gauge group is $\mathcal{T}_{uv}$ with a shorter periodicity for $\te$ and hence with larger axial symmetry. See figure \ref{flow}. 
The outline for the rest of the paper is as follows.
In section \ref{sec:reviewQCD} we will review some known results about $N_f\geq 2$ QCD which will be useful for later.
In section \ref{sec:Nf1qcd} we will study the simplest example where the prescription described here can be used, which is $N_f=1$ QCD. In this example we will show how the new anomaly constraints can be used to re-derive some known results about $N_f=1$ QCD. These results include the existence of a cusp in the effective potential for the $\eta'$ meson with a certain Chern-Simons theory living on it, and the construction of baryons from the low energy perspective. Then we move on to study a family of chiral gauge theories in sections \ref{sec:AS}-\ref{sec:sym}. For these theories, the anomaly has two interesting new applications. The first regards large $N$. The anomaly implies that a certain operator $\mathcal{B}$ must condense in the IR. This operator is made out of $O(2N)$ fermions and its condensation stands in contradiction to the large $N$ arguments of \cite{Eichten:1985fs}. The anomaly constraints and the large $N$ arguments together imply that there is no consistent confining phase for these theories. We argue that the large $N$ arguments made in \cite{Eichten:1985fs} are actually not completely correct. 

 Another application is to use the anomaly to predict phases of multi-flavour versions of some theories. As we show, the condensate of the operator $\mathcal{B}$ doesn't break any symmetry in the one-flavour theories but does break in their multi-flavour generalizations. This condensation which is predicted by the new anomaly, provides an easy and natural way to give new proposals for the IR phases of the multi-flavour theories. While the one-flavour and multi-flavour theories look very different with different patterns of symmetry breaking, our proposal shows that they are all controlled by the same condensate. In section \ref{sec:2gaugegroups} we give more examples in which the anomaly predicts certain condensates in some one-flavour theories. This condensate can be used to predict new phases for the multi-flavour generalizations of these theories. Some technical details are discussed in the appendices. 
\section{$SU(N)$ Yang-Mills with $\te\in[0,2\pi/N]$}
\label{sec:YM}
In this section we will describe the mechanism that leads to a YM theory with fractional periodicity for $\te$. Consider $SU(N_1)\times SU(N_2)$ gauge theory with one massless bifundamental Dirac fermion. We will denote this fermion by $\la_D$. We also define $L=gcd(N_1,N_2)$ and $L_{1,2}=\frac{N_{1,2}}{L}$. The uv Lagrangian is parametrized by two gauge couplings, and one theta angle. The reason is that one theta angle becomes unphysical due to the ABJ anomaly. Under axial rotations of $\la_D$, the theta angles are shifted by
\eql{axialla}{\la_D\to e^{i\gamma_5\be}\la_D\ \Rightarrow\ \te_1\to\te_1+2N_2\be\ ,\ \te_2\to\te_2+2N_1\be\ .} 
The linear combination $\te_p=L_1\te_1-L_2\te_2$ is invariant under axial rotations and hence physical. The orthogonal linear combination can always be taken to zero using axial rotations and hence unphysical.
The symmetries of this model are
\begin{itemize}
	\item $\mathbb{Z}_L$ 1-form symmetry: This is a subgroup of the $SU(N_1)\times SU(N_2)$ center that leaves the fermion invariant.
	\item $U(1)_\la$: $\la_D\to e^{i\al}\la_D$.
	\item $\mathbb{Z}_L$ axial symmetry: It can be seen from \eqref{axialla} that for $\be=\frac{\pi}{L}\mathbb{Z}$, the two theta angles are shifted by $2\pi\mathbb{Z}$, hence it is a symmetry. \footnote{Notice that $\be=\pi$ is gauge equivalent to the identity if either $N_1$ or $N_2$ is even. If both $N_{1,2}$ are odd, then $\be=\pi$ is part of $U(1)_\la$. In any case, there is only a $\mathbb{Z}_L$ transformation which is truly axial.}
	\item $T$: There is a time reversal symmetry whenever $\te_p=\pi\mathbb{Z}$. 
\end{itemize}
We can analyze the theory in the limit where $SU(N_2)$ confines when $SU(N_1)$ is still weakly coupled. The effective theory after the confinement of $SU(N_2)$ is an $SU(N_1)$ gauge theory coupled weakly to an $SU(N_1)$ NLSM parametrized by the matrix $U\in SU(N_1)$. The interactions with the $SU(N_1)$ gauge fields generate a potential on the target space \cite{Karasik:2019bxn}. This happens in two steps. First, perturbatively, the target space is lifted except for the central elements $U=e^{2\pi i k/N_1}$ with $k\in\mathbb{Z}$. There are $N_1$ minima labelled by an integer $k$. Around each minimum, the effective theory is an $SU(N_1)$ pure YM with the theta angle, \eq{\te^{(k)}=\onov{L_1}\te_p+\frac{2\pi L_2}{L_1}k\ .}
The $SU(N_1)$ YM theory then confines as usual, and the energy at each one of the $k$-vacua depends on $\te^{(k)}\ mod\ 2\pi$. The true vacua are those with minimal $|\te^{(k)}|$. We see that $\te^{(k)}=\te^{(k+L_1)}\ mod\ 2\pi$. The effective theta angle splits the $N_1$ minima into $L_1$ sets of $L$ degenerate minima, related by the action of the axial $\mathbb{Z}_L$. For generic $\te_p$, there is a $\mathbb{Z}_L$ degeneracy due to the spontaneous breaking of $\mathbb{Z}_L$. For $\te_p=\pi+2\pi\mathbb{Z}$, there is a $2L$ degeneracy due to an additional breaking of $T$.
We can work in the limit where $\Lambda_2\to\infty$ (the strong coupling scale of $SU(N_2)$). When we sit in one vacuum, the effective theory we get is locally identical to $SU(N_1)$ YM but with subtle modifications. First, the periodicity of $\te$ is not $2\pi$ but $\frac{2\pi}{L_1}$. Similarly, time reversal is a symmetry when $\te=\frac{\pi}{L_1}\mathbb{Z}$. To see this, consider starting from $\te=0$ and increasing $\te$ continuously. As we cross $\te=\frac{\pi}{L_1}$ we jump from the $k=0\mod L_1$ vacua to the $k=-1\mod{L_1}$ vacua. This reflects the first order phase transition that happens usually at $\te=\pi$. To get the shortest effective periodicity for $\te$, we will choose $N_2$ such that $L_1=N_1$ (and therefore $L=1$). For example, we can take $N_2=N_1-1$. In this case we have a unique vacuum for generic $\te$, and twofold degeneracy for $\te=\frac{\pi}{N_1}$. The theory is  $SU(N_1)$ YM with $\frac{2\pi}{N_1}$ periodicity of $\te$.
 What about 1-form symmetries? in the full theory, there is only a $\mathbb{Z}_L$ symmetry, which for $L=1$ means there is no 1-form symmetry. However, the completion to $\mathbb{Z}_{N_1}$ 1-form symmetry emerges at energies $E\ll\Lambda_2$. Another way to say it is that screening by $\la_D$ will not occur at energies $E\ll\Lambda_2$, and hence the fundamental Wilson line of $SU(N_1)$ becomes stable. 
 In the next sections we will discuss some applications of this theory. The main idea is as follows: Consider some 4d gauge theory with gauge group $SU(N)$ and some matter content, which is our theory of interest. Assume that this theory has no non-trivial discrete axial transformations (if it does, the constraints that we get will not teach us anything new). We can imagine adding at higher energies $SU(N-1)$ gauge theory and a massless bifundamental Dirac fermion $\la_D$. This effectively changes the periodicity of $\te$ from $2\pi$ to $\frac{2\pi}{N}$. Using this shorter periodicity, we can derive new constraints on the flow of the theory. 
\section{Review of some aspects of $N_f\geq 2$ QCD}
\label{sec:reviewQCD}
The simplest example in which we can use our new anomaly constraints is $N_f=1$ QCD. The constraints can be used to derive rigorously some known results about $N_f=1$ QCD in which it behaves exactly as expected. In this section we will review some known results about $N_f\geq 2$ QCD from a perspective that will make the discussion about $N_f=1$ QCD in section \ref{sec:Nf1qcd} clearer. 
Our starting point is 4d $SU(N)$ gauge theory with $N_f\geq 2$ fundamental Dirac fermions $\psi_D$. In the massless limit, the theory enjoys the symmetry $SU(N_f)_L\times SU(N_f)_R\times U(1)_V$ where $SU(N_f)_{L,R}$ acts only on the left/right handed fermions, and $U(1)_V$ acts in a vector-like fashion $\psi_D\to e^{i\al}\psi_D$. $U(1)_A$ transformations $\psi_D\to e^{i\al\gamma_5}\psi_D$ shift the $\te$ term by $\te\to\te+2N_f\al$. This breaks $U(1)_A\to \Z_{2N_f}$. However, this $\Z_{2N_f}$ is embedded inside $SU(N_f)_L\times SU(N_f)_R\times U(1)$ (see for example equation (2.11) of \cite{Tanizaki:2018wtg}) so we don't need to consider it as an independent symmetry.

These symmetries don't act faithfully on gauge invariant operators, we will go back to this point later. 't-Hooft anomalies cannot be matched by massless gauge invariant composite fermions \cite{tHooft:1979rat} and assuming confinement, the chiral symmetry must be broken. The symmetry breaking pattern is \eq{SU(N_f)_L\times SU(N_f)_R\times U(1)_V\to SU(N_f)_V\times U(1)_V\ .}
The breaking is due to the chiral condensate $\vev{\overline{\psi}_D\psi_D}_i^j\sim \delta_i^j$ where $i,j$ are flavour indices. The low energy theory is an $SU(N_f)$ non-linear sigma model (NLSM) parametrized by $U\in SU(N_f)$. Many of the anomalies such as $SU(N_f)_L^3,\ SU(N_f)_R^3$ are matched by the Wess-Zumino (WZ) term,\cite{Wess:1971yu,Witten:1983tw}
\eq{S_{WZ}=-\frac{iN}{240\pi^2}tr\int_{\mathcal{M}_5}(dUU^\dagger)^5\ .}
Here $\mathcal{M}_5$ is a 5d manifold whose boundary is the 4d manifold on which the theory is defined. 
However, we want to mention here two anomalies that are matched in the IR not by the WZ term.
\subsection{$\te$ periodicity anomaly}
\label{subsec:theta}
We can deform the theory in the uv by giving the quarks a small mass $M=me^{i\te/N_f}$ with $m\ll\Lambda$ where $\Lambda$ is the strong coupling scale. Here we absorbed the $\te$-term into the mass term of the fermions. The mass term breaks explicitly the chiral symmetry down to the vector-like symmetries $SU(N_f)_V\times U(1)_V$. This symmetry doesn't act faithfully on gauge invariant operators. The correct faithfull symmetry group is \eq{G=\frac{U(N_f)_V}{\Z_N}=\frac{SU(N_f)_V\times U(1)_V}{\Z_{N_f}\times \Z_N}\ .} 
When $gcd(N,N_f)\neq 1$, there is a mixed anomaly between the $2\pi$ periodicity of $\te$ and the symmetry $G$ \cite{Cordova:2019uob}. If we couple $G$ to background gauge fields, the theory stops being periodic under $\te\to\te+2\pi$. This can be shown as follows. Define $a$ as the dynamical $SU(N)$ gauge fields, and $f$ as the field strength. We will introduce the background $SU(N_f)$ and $U(1)$ gauge fields denoted by $A_f$ and $A_V$ respectively. To quotient by $\Z_N\times \Z_{N_f}$ we will introduce two pairs of $U(1)$ 1-form and 2-form gauge fields $(B_c^{(1)},B_c^{(2)})$ and $(B_f^{(1)},B_f^{(2)})$ subject to the constraints \eq{dB_c^{(1)}=NB_c^{(2)}\ ,\ dB_f^{(1)}=N_fB_f^{(2)}\ ,} with the 1-form gauge redundancies \eql{1formgauge}{&B_c^{(1)}\to B_c^{(1)}+N\omega_c^{(1)}\ ,\ B_c^{(2)}\to B_c^{(2)}+d\omega_c^{(1)}\ ,\\& B_f^{(1)}\to B_f^{(1)}+N_f\omega_f^{(1)}\ ,\ B_f^{(2)}\to B_f^{(2)}+d\omega_f^{(1)}\ .}
Here $\omega_{c,f}^{(1)}$ are two arbitrary 1-forms parametrizing the gauge transformation.
We promote the $SU(N)$ gauge fields to $U(N)$ gauge fields $a\to\tilde{a}=a+\onov{N}B_c^{(1)}\boldmath{1}_N$ with the $U(N)$ field strength $\tilde{f}$. Similarly, we promote the $SU(N_f)$ gauge fields $A_f$ to $\tilde{A}_f=A_f+\onov{N_f}B_f^{(1)}\boldmath{1}_{N_f}$ with the field strength $\tilde{F}_A$. $\tilde{f}$ and $\tilde{F}_A$ are not invariant under \eqref{1formgauge} and we should replace them with the combinations $\tilde{f}-B_c^{(2)}\boldmath{1}_N$ and $\tilde{F}_A-B_f^{(2)}\boldmath{1}_{N_f}$. Finally, we impose the 1-form gauge transformations on the $U(1)$ gauge field
\eq{A_V\to A_V-\omega_c^{(1)}-\omega_f^{(1)}\ .}
The associated field strength is the gauge invariant combination
\eq{F_V=dA_V+B_c^{(2)}+B_f^{(2)}\ .}

 Under $2\pi$ shifts of $\te$, the action is shifted by 
\eq{\delta S_\te=\frac{2\pi}{8\pi^2}tr\int \left(\tilde{f}-B_c^{(2)}\boldmath{1}_N\right)^2=\frac{2\pi}{8\pi^2}tr\int \left(\tilde{f}^2\right)-\frac{2\pi N}{8\pi^2}\int \left(B_c^{(2)}\right)^2\ ,}
where $\frac{1}{8\pi^2}tr\int \left(\tilde{f}^2\right)\in \Z$ but $\frac{ N}{8\pi^2}\int \left(B_c^{(2)}\right)^2\in \onov{N}\Z$. We see that the periodicity of $\te$ is violated. However, we have the freedom to add counterterms of the form
\eq{S_{ct}=\frac{c_1\te}{8\pi^2}tr\int \left( \tilde{F}_A-B_f^{(2)}\boldmath{1}_{N_f}\right)^2+\frac{c_2\te}{8\pi^2}\int F_V^2\ ,}
with arbitrary coefficients $c_{1,2}$. If we choose $c_1=Nk\ ,\ c_2=NN_fk$ with $k\in\mathbb{Z}$, the action is shifted by
\eq{\delta(S_\te+S_{ct})=-\frac{2\pi N(1-N_fk)}{8\pi^2}\int \left(B_c^{(2)}\right)^2 \mod\ 2\pi\Z\ .}
The periodicity can be restored if there exists a solution to $1-N_fk\in N\Z$ which is true only if $gcd(N,N_f)=1$. For generic $N,N_f$ the periodicity is violated. A convenient way to phrase this anomaly is through the constraint on the partition function,
\eql{thetaanomaly}{&\mathcal{Z}[\te=2\pi]=e^{iS_{A}}\mathcal{Z}[\te=0]\ ,\\
&S_{A}=\frac{N}{4\pi N_f}\tr\int \left(\tilde{F}_A+dA_V+B_c^{(2)}\right)^2-\frac{N}{4\pi}\int \left(B_c^{(2)}\right)^2\mod S_{ct}\ ,}
where for later convenience we chose to represent the anomaly with the specific choice of $c_1=\frac{N}{N_f},\ c_2=N$. Next we will review how this constraint is satisfied in the IR.

In the IR, the mass term lifts the $SU(N_f)$ target space of the NLSM due to the potential
\eq{V_m\sim \Lambda^3 tr (MU+M^*U^\dagger)+O(\Lambda^2m^2)\ .}
By taking an ansatz $U_k=e^{2\pi ik/N_f}$, we get
\eq{V_m(U_k)\sim \Lambda^3m \cos\left(\frac{2\pi k+\te}{N_f}\right)\ .}
For generic $\te$ we find a unique gapped vacuum. In particular, for $\te=0$ the vacuum is $U=1$ while for $\te=2\pi$ the vacuum is $U=e^{-2\pi i/N_f}$. The anomaly constraint \eqref{thetaanomaly} implies that any trajectory in field space connecting the two vacua must carry the anomaly. Explicitly, we can consider a setup in which we put the theory on $\mathcal{M}_5=\mathcal{M}_4\times I$ where $I$ is an interval parametrized by $y\in[0,1]$. Along the interval we change $\te$ continuously from $0$ to $2\pi$ such that the theory goes from $U=1$ at $y=0$ to $U=e^{-2\pi i/N_f}$ $y=1$. The action of the theory must satisfy
\eq{S_{A}=\int_0^1 dy \frac{d}{dy}S_\gamma \mod 2\pi\Z\ ,}
where $S_\gamma$ is the action along any trajectory in field space $\gamma$ with the boundary conditions stated above.

 The simplest trajectory we can consider is through the $SU(N_f)$ matrix. Indeed, it was shown in \cite{Gaiotto:2017tne} that this trajectory captures the anomaly correctly. For this trajectory, the contribution to the anomaly comes from the WZ term. Another trajectory which is under control at least in the large $N$ limit is by changing $\eta'$. This can be easily seen from the coupling of $\eta'$ to the background $U(N_f)/\Z_N$ gauge fields as fixed from the gauged WZ term in the large $N$ limit. We will review how it works, why it implies that there is a cusp and what it tells us about the cusp. 
We will start by coupling the theory to $U(N_f)$ gauge fields $\tilde{A}_f$ and $A_V$ without the quotient by $\Z_N$. This is equivalent to setting $B_c^{(1,2)}=0$ in the equations above. In the large $N$ limit, we promote the target space $U\in SU(N_f)\to U\in U(N_f)$ where $det\ U=e^{i\eta'}$. The part in the gauged WZ action that contains the $\eta'$ field is
\eql{Seta}{S_{\eta'}&=-\frac{N}{8\pi^2N_f}\int_{\mathcal{M}_5}\tr (F^2)d\eta'+\frac{iN}{48\pi^2N_f}\int_{\mathcal{M}_5} d \tr(FDUU^\dagger+FDU^\dagger U)d\eta'\ ,}
where $F=\tilde{F}_A+dA_V$ is a $U(N_f)$ field strength and $DU=dU-i\tilde{A}_fU+iU\tilde{A}_f$ is the covariant derivative of $U$. We consider the trajectory in which we change continuously $\eta'$ from $0$ to $-2\pi$ along the interval $I$. The contribution from \eqref{Seta} is 
\eql{SAeta}{S_{A,\eta'}=\frac{N}{4\pi N_f}\int_{\mathcal{M}_4}\tr (F^2)\ ,}
which is equivalent to \eqref{thetaanomaly} when setting $B_c^{(1,2)}=0$.
Notice that \eq{N_fS_{A,\eta'}=\frac{N}{4\pi}\int_{\mathcal{M}_4}tr(F^2)\in 2\pi\mathbb{Z}\ .}
This reflects the fact that $\eta'=0$ and $\eta'=2\pi N_f$ are the same point $U=1$ and the anomaly between them must vanish. Another way to view it is to see that under constant shifts $\eta'\to \eta'+c$, \eqref{Seta} is shifted by\footnote{The second term in \eqref{Seta} is of the form $\int dK_3\wedge d\eta'$ where $K_3$ is a well defined gauge invariant 3-form. Therefore we can write it as a boundary term $\int K_3\wedge d\eta'$ which is invariant under constant $\eta'$ shifts.}
\eq{\delta S_{\eta'}=-\frac{cN}{8\pi^2N_f}\int_{\partial\mathcal{M}_5}tr(F^2)\ .}
We see that our theory is single-valued under $\eta'\to\eta'+2\pi N_f$ as should be from its definition. However this raises a question. \eqref{Seta} can be written as a well defined gauge invariant 4d action,
\eq{S_{\eta'}&=-\frac{N}{8\pi^2N_f}\int_{\partial\mathcal{M}_5}\tr (F^2)\eta'+\frac{iN}{48\pi^2N_f}\int_{\partial\mathcal{M}_5}  \tr(FDUU^\dagger+FDU^\dagger U)d\eta'\ .}
In the context of anomalies, one would say that this is a counterterm. Such a term cannot carry an anomaly. This is very different from the other parts of the gauged WZ action. The terms that don't involve $\eta'$ must be added to the theory as part of the gauging procedure since the theory is not well defined without them. $\eta'$ on the other hand doesn't appear in the WZ term and the theory is well defined without \eqref{Seta}. What happens when we quotient by the $\Z_N$ and bring back to life $B_c^{(1,2)}$? The anomaly action \eqref{SAeta} then becomes
\eq{S_{A,\eta'}=\frac{N}{4\pi N_f}\int_{\mathcal{M}_4}tr\left(F+B_c^{(2)}\right)^2\  .}
Here it looks like the situation is even worse. It is not equal to \eqref{thetaanomaly} and it is not invariant under $\eta'\to\eta'+2\pi N_f$ which makes the theory ill-defined. The missing ingredient is of course the cusp at $\eta'=\pi$. The full theory, i.e. \eqref{Seta} plus the theory on the cusp together must be single valued under $\eta'\to\eta'+2\pi N_f$. Therefore, the theory on the cusp must be such that as you couple it to background $U(N_f)/\Z_N$ gauge fields, the action is shifted by the bulk term \eq{\delta S_{cusp}=-\frac{N}{4\pi}\int_{\mathcal{M}_4}(B^{(2)})^2\ mod\ \frac{2\pi}{N_f}\ ,} such that the full theory is invariant under $\eta'\to \eta'+2\pi N_f$.

 This is exactly what you get if the theory on the cusp is for example $SU(N)_{1}$ CS theory \cite{Gaiotto:2017tne}. $SU(N)_{1}$ is a well defined 3d theory. However, if we couple to $\Z_N$ gauge fields, the CS theory itself becomes ill-defined. This tells us that we must add something in the bulk as part of the gauging procedure to make the full theory well defined. \eqref{Seta} does the job and together with the cusp carries the anomaly. Following this point of view, the addition of \eqref{Seta} to the theory has nothing to do with the gauged WZ action. \eqref{Seta} needs to be added to the theory to make the cusp well defined while the gauged WZ action is perfectly well defined without \eqref{Seta}. In the large $N$ limit, we can derive \eqref{Seta} from the gauged WZ action due to dynamical reasons but its kinematic origin remains the cusp.

We will finish this section with a comment about finite $N$. At finite $N$ we cannot separate $\eta'$ from other massive excitations and it makes no sense to talk about it by itself. Let us define \eql{Y}{Y=\vev{\overline{\psi}\psi}^{-N_f}det_{N_f}(\overline{\psi}\psi)\ ,} and consider the potential $V(Y)$ in the complex $Y$ plane. The potential has a minimum point at $Y=1$. At large $N$, the ring $|Y|=1$ looks like a valley which reflects the fact that $\eta'$ is a light particle. It is only then that we can consider motion along this ring while ignoring other massive excitations. This allows us to write its low energy effective action (together with the pions) and derive the existence of the cusp. However, the anomaly matching conditions must hold also for small $N$ when $\eta'$ cannot be defined. What the anomaly tells us in general is that it should be impossible to form a closed loop on the $|Y|$ plane winding $Y=0$ while remaining trivially gapped. Let's assume that such a loop exists and denote the coordinate on the loop by $t\in[0,2\pi]$. Any coupling of $t$ to the background gauge fields must be such that it is unchanged under $t\to t+2\pi N_f$. Therefore, any coupling of $t$ to the background gauge fields is nothing but a local counterterm and cannot carry the anomaly. Hence, the anomaly matching conditions are violated. The simplest resolution is that there is a branch cut at $arg(Y)=\pi$ at least for large enough $|Y|$. For small $|Y|$ the theory might be for example in a deconfining phase and the anomaly can be matched in other ways such as gapless excitations. For $N_f=1$ QCD, there is no anomaly of this type. In section \ref{sec:Nf1qcd} we will argue that similar constraints can be rigorously derived using the method explained in the introduction.
\subsection{$U(1)_B-SU(N_f)^2_{L,R}$ anomaly}
\label{subsec:skyrme}
The second anomaly we want to mention is the mixed triangle anomaly between $U(1)_B$ and $SU(N_f)_L^2$ or $SU(N_f)_R^2$. Similar to the previous anomaly, also here in the large $N$ limit this anomaly can be derived from the WZ term as was shown by Witten \cite{Witten:1983tw}. The reason is that in the large $N$ limit the target space is promoted from $U\in SU(N_f)$ to $U\in U(N_f)$. Consider a general vector-like $U(1)_Q$ transformation $U\to e^{-iQ\al}U e^{iQ\al}$ where $Q$ is the matrix of charges. This $U(1)_Q$ is part of $SU(N_f)_L\times SU(N_f)_R$ and therefore has mixed anomalies with $SU(N_f)_{L,R}^2$ as part of the $SU(N_f)_{L,R}^3$ anomalies. For $U\in U(N_f)$, we can have $tr(Q)\neq 1$. This allows us to take continuously the limit $Q\to \onov{N}$ which takes $U(1)_Q\to U(1)_B$ and the $U(1)_Q\times SU(N_f)_{L,R}^2$ anomalies to $U(1)_B\times SU(N_f)_{L,R}^2$ anomalies. In this way, the $U(1)_B\times SU(N_f)_{L,R}^2$ anomalies can be obtained from the WZ term.
However, this derivation cannot be the entire story. The simplest way to see that something goes wrong with it is to consider $N_f=2$. In this case there is no WZ term and no (continuous) $SU(N_f)_{L,R}^3$ anomalies. Yet, the $U(1)_B\times SU(N_f)_{L,R}^2$ anomaly still exists. In this case, the $U(1)_B\times SU(N_f)_{L,R}^2$ anomaly doesn't come from the WZ term and must have another source. The resolution is of course skyrmions. In $SU(N_f)$ NLSM, there is a topological conserved current, also known as the skyrmion current $S=\onov{24\pi^2}tr(dUU^\dagger)^3$. The topological $U(1)_S$ associated with this current has mixed anomalies with $SU(N_f)_{L,R}^2$. These anomalies exist independently of the WZ term. It doesn't matter what the level of the WZ term is or even if it exists at all, the $U(1)_S\times SU(N_f)_{L,R}^2$ anomalies are the same. To see it, couple to $SU(N_f)_L$ background gauge fields $A_L$. The covariant derivative of $U$ becomes $DUU^\dagger =dUU^\dagger-iA_L$. One can show that it is impossible to write a conserved gauge invariant current for $U(1)_S$ which means that if we gauge $SU(N_f)_L$, $U(1)_S$ is broken. Hence, there is a mixed anomaly between them. The best that we can achieve is to define the current \eq{S'=\onov{24\pi^2}tr[(DUU^\dagger)^3+3iFDUU^\dagger]\ .}
$S'$ is not conserved, but satisfies $dS'=\onov{8\pi^2}tr(F_L^2)$ which is the conventional form of the anomaly.
This anomaly is one of the strongest proofs in the identification of $U(1)_S$ as the low energy description of $U(1)_B$. An interesting consequence of this anomaly is that if we turn on $A_L$ such that on some 3-manifold, $\onov{8\pi^2}\int tr\left(A_L\wedge dA_L-\frac{2i}{3}A_L^3\right)=-1$, the vacuum of this configuration will have baryon charge 1. This is satisfied because the vacuum equation is $dUU^\dagger=iA_L$. This immediately gives
\eq{\onov{24\pi^2}\int tr\left(dUU^\dagger\right)^3=1\ ,}
in agreement with the identification of the skyrmion current and the baryon current.

\section{$N_f=1$ QCD}
\label{sec:Nf1qcd}
The simplest example in which we can make use of the method described in the introduction is $N_f=1$ QCD. The theory is defined as a 4d $SU(N)$ gauge theory coupled to one massless Dirac fermion, $\psi_D$. The only symmetry of this theory is the baryon symmetry $U(1)_\psi$ that acts as $\psi_D\to e^{i\al_\psi}\psi_D$. $U(1)_\psi$ is anomaly free and $gcd(N,N_f)=1$. Therefore the theory is completely free of any anomalies. This case is very different from the $N_f\geq 2$ case where anomalies highly constrain the low energy physics as reviewed in \ref{sec:reviewQCD}. Assuming the theory is confining and gapped at low energies, we can make the following claims regarding $N_f=1$ QCD:
\begin{enumerate}
	\item The fermion bilinear $\overline{\psi}_D\psi_D$ condenses at low energies.
	\item Assuming $\vev{\overline{\psi}_D\psi_D}\neq 0$, large $N$ arguments \cite{Veneziano:1979ec,Witten:1980sp,DiVecchia:1980yfw} imply that there is a cusp at $\eta'=\pi$. The existence of a cusp holds also at finite $N$. Similar to the discussion around equation \eqref{Y}, this can be rephrased as a statement about the potential of $Y=\onov{\vev{\overline{\psi}_D\psi_D}}\overline{\psi}_D\psi_D$ since $\eta'$ is not well defined as a periodic scalar at finite $N$.
	\item The 3d effective theory living on the cusp is constrained by anomalies of $\mathcal{T}_{x-ray}$. A possible candidate is an $SU(N)_{1}$ CS theory (possibly coupled to some matter as in \cite{Gaiotto:2017tne}). This fixes the coupling of $\eta'$ to a background $U(1)_\psi$ gauge field.
	\item The low energy description of baryons in $N_f=1$ QCD is also constrained by anomalies of $\mathcal{T}_{x-ray}$. The constraints are consistent with the construction of baryons that was described in \cite{Komargodski:2018odf, Karasik:2020pwu,Karasik:2020zyo}.
\end{enumerate}
The conclusions of this analysis are, not surprisingly, in exact agreement with what one would expect from large $N$ or higher $N_f$. Yet, until now, there was no kinematic argument explaining why all this should hold. In this case, the new anomaly can give a rigorous derivation of things we already know (or at least strongly believe) that are true.
\subsection{$\mathcal{T}_{x-ray}$: Kinematics}
In this section we will embed $N_f=1$ QCD inside a bigger theory $\mathcal{T}_{x-ray}$ as explained in the introduction. We will study the constraints of $\mathcal{T}_{x-ray}$ and show how they lead to constraints on $N_f=1$ QCD. 
Consider starting from an $SU(N)\times SU(M)$ gauge theory with one Dirac fermion $\psi_D$ in the $(\Box,1)$, and one Dirac fermion, $\la_D$ in the $(\Box,\overline{\Box})$.  First we will discuss the global symmetries of the theory. There are the two vector-like transformations $U(1)_\psi\times U(1)_\la$ which we take to act as \eq{U(1)_\psi:\ \psi_D\to e^{i\al_\psi}\psi_D\ ,\ \la_D\to e^{i\al_\psi}\la_D\ ,\ U(1)_\la:\ \psi_D\to \psi_D\ ,\ \la_D\to e^{i\al_\la}\la_D\ .}
It is manifest that $\al_\psi=\frac{2\pi}{N}$ and similarly $\al_\la=\frac{2\pi}{M}$ are gauge equivalent to the identity. 
Under general transformations that act only on the left handed fermions,
\eq{\psi_L\to e^{ i \omega_\psi }\psi_L\ ,\ \la_L\to e^{i \omega_\la}\la_L\ ,}
the action is shifted by
\eq{\delta S=-\omega_\psi q_N- M\omega_\la q_N- N\omega_\la q_M\ .}
Here $q_{N,M}=\onov{8\pi^2}\int tr(f_{N,M}\wedge f_{N,M})\in\mathbb{Z}$ are the topological densities for $SU(N)$ and $SU(M)$ respectively.
This transformation is a symmetry if 
\eq{\omega_\la=\frac{2\pi k}{N}\ ,\ \omega_\psi=2\pi m-\frac{2M\pi k}{N}\ ,}
with $k,m\in\Z$. To maximize the axial transformations of $\psi_L$, we will choose $M$ such that $gcd(N,M)=1$. 
For simplicity and concreteness, we will take $M=N-1$ such that the axial transformations with\eq{\omega_\psi=\omega_\la=\frac{2\pi k}{N}\ ,} generate an exact $(\Z_N)_L$ symmetry.
What are the constraints that we get from $(\Z_N)_L$? Consider coupling the theory to $U(1)_\psi\times U(1)_\la$ background gauge fields. This is done by promoting the $SU(N),\ SU(M)$ gauge fields $a_{N,M}$ to $U(N),\ U(M)$ gauge fields by defining
\eq{\tilde{a}_N=a_N+\onov{N}A_\psi\ ,\ \tilde{a}_{N-1}=a_{N-1}+\onov{N-1}A_\la\ .}
As shown in \ref{sec:detanomaly}, there is a $(\Z_N)_L\times U(1)_\psi\times U(1)_\la$ mixed anomaly. Under $(\Z_N)_L$, the partition function transforms as
\eql{anomalycondition}{\mathcal{Z}\to exp\left(\frac{i}{2\pi N}\int dA_\psi\wedge dA_\la\right)\mathcal{Z}\ ,\ \frac{1}{2\pi N}\int dA_\psi\wedge dA_\la\in \frac{2\pi}{N}\Z\ .} 
This property must hold also in the IR. In the next section we will follow the flow of the theory to see explicitly the implication of the anomaly. 

\subsection{Dynamics and the cusp}
\label{subsec:cusp}
To study the dynamics of $\mathcal{T}_{x-ray}$, we will take the limit in which $\Lambda_{N-1}\gg\Lambda_N$ such that $SU(N-1)$ confines when $SU(N)$ is still weakly coupled. From the point of view of $SU(N-1)$, $\la_D$ looks like $N$ fundamental Dirac fermions. At energies $\Lambda_N\ll E\ll\Lambda_{N-1}$, the confinement of $SU(N-1)$ results in a $\overline{\la}_D\la_D$ condensate and an $SU(N)$ NLSM parametrized by the matrix $U_\la\in SU(N)$. In addition we still have the $SU(N)$ gauge fields and the fundamental fermion $\psi_D$. The Sigma model matrix transforms under $SU(N)$ gauge transformations $V\in SU(N)$ as $U\to VU_\la V^\dagger$. The condition \eqref{anomalycondition} is satisfied in this regime in the following way. $(\Z_N)_L$ transformations act as $\psi_L\to e^{2\pi i/N}\psi_L$ as before together with rotation of the $SU(N)$ matrix, $U_\la\to e^{2\pi i/N}U_\la$. As shown in \ref{sec:detanomaly}, the rotation of $\psi_L$ changes the action by $\delta S_\psi$ and the rotation of $U_\la$ changes the action by $\delta S_\la$ which are specified in \eqref{deltaSs}. The two contributions together $\delta S_{tot}=\delta S_\psi+\delta S_\la$ satisfy the anomaly matching condition \eqref{anomalycondition}.
Now we can continue flowing down. First, interactions of $U_\la$ with the $SU(N)$ gauge fields will generate a potential on the target space. As a result, only the center elements $U_\la=e^{2\pi ik/N}\boldmath{1}_N$ remain the vacua of the theory \cite{Karasik:2019bxn}. The choice of vacuum breaks spontaneously the $\Z_N$ axial symmetry. The mass of the sigma model fluctuations around the vacuum is proportional to $\Lambda_{N-1}$ which we take to be as large as we want such that they can be completely ignored. Sitting in one specific vacuum, the effective theory is exactly $N_f=1$ QCD made out of the $SU(N)$ gauge fields and the fermion $\psi_D$. The dynamics doesn't know about the existence of the sigma model and the other discrete vacua. The flow from this point is equivalent to the flow of $N_f=1$ QCD, and the IR phase we get here is equivalent to the IR phase of $N_f=1$ QCD. This implies that $N_f=1$ QCD must satisfy the anomaly conditions coming from the full x-ray theory.

\begin{figure}
	\vspace{4pt}
	\begin{center}
		\includegraphics[width=0.9\textwidth]{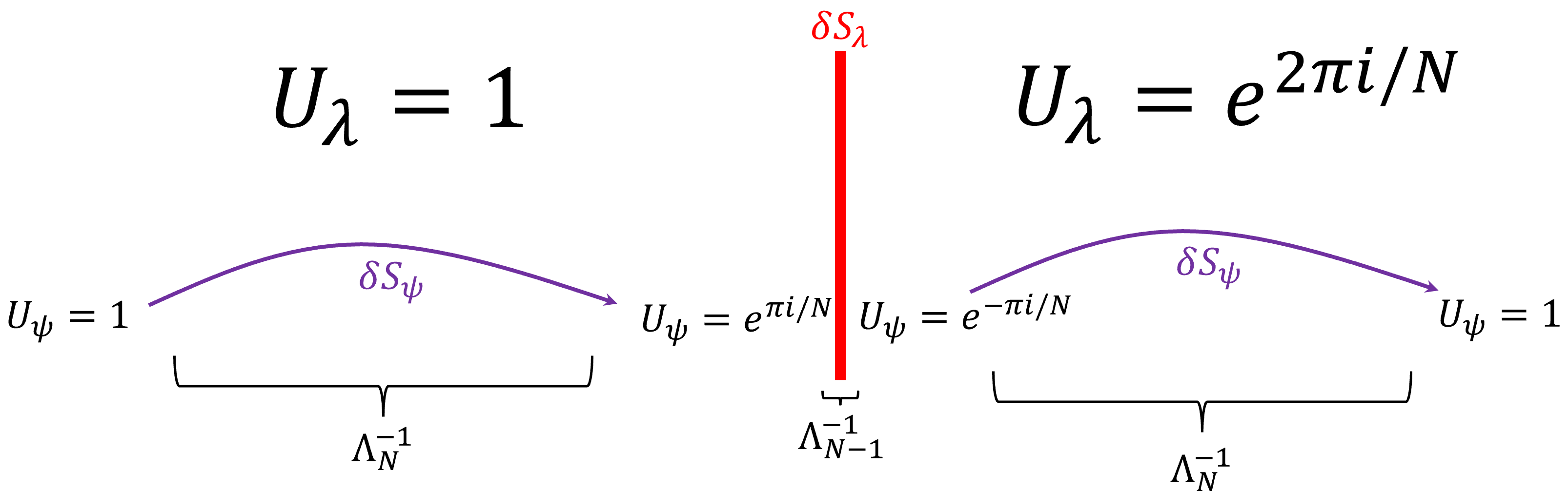}
		\caption{\footnotesize{An illustration of a domain wall configuration that interpolates between the vacuum $U_\la=1\ ,\ U_\psi=1$ on the left to $U_\la=e^{2\pi i/N},\ U_\psi=1$ on the right. Due to the scale separation $\Lambda_{N-1}\gg\Lambda_N$, the DW is described as follows. The phase of $U_\psi$ changes along an interval of length $\sim\Lambda_N^{-1}$ from $0$ to $\pi/N$. At this point there is a rapid jump (at length $\Lambda_{N-1}^{-1}$) between the two $U_\la$ vacua }}\label{DW}
	\end{center}
	\vspace{4pt}
\end{figure}

So what happens in the IR? If $N_f=1$ QCD confines without a condensate $\vev{\overline{\psi}_D\psi_D}=0$, $(\Z_N)_L$ transformations will change the action only by $S_\la$ which doesn't satisfy \eqref{anomalycondition} by itself. Therefore, this phase is inconsistent. The obvious resolution is that $\vev{\overline{\psi}_D\psi_D}\neq 0$ also for $N_f=1$ QCD and the phase of the condensate carries the anomaly. To see how it works we consider a domain wall configuration as in figure \ref{DW}

The consequences of this constraint are easily phrased in the large $N$ limit where the effective theory for $\eta'$ is well defined. The coupling of $\eta'$ to the $U(1)_\psi$ gauge field is fixed by \eqref{anomalycondition} and \eqref{deltaSs} to be,
\eql{etaA}{\lag_{\eta, A_\psi}=-\onov{8\pi^2N}\eta' dA_\psi\wedge dA_\psi\ .}
Similar to the discussion in section \ref{sec:reviewQCD}, also here we must demand that the theory is single valued under $\eta'\to\eta'+2\pi$. This is violated by \eqref{etaA} which means that something must exist to restore the periodicity. The simplest resolution is the existence of a cusp at $\eta'=\pi$. The effective theory that lives on the cusp must be such that when we couple to $A_\psi$ and add the term \eqref{etaA}, the theory is single-valued under $\eta'\to\eta'+2\pi$. This happens if the theory on the cusp is for example $SU(N)_{1}$ Chern-Simons theory. The dynamical CS theory is well defined because on any closed 4d manifold $\onov{8\pi^2}tr(f\wedge f)\in\Z$.  
When we couple to $A_\psi$, the dynamical CS field $a$ is combined with $A_\psi$ to form a $U(N)$ gauge field. Together they satisfy \eql{UNquant}{\onov{8\pi^2}tr(f\wedge f)+\onov{8\pi^2N}dA_\psi\wedge dA_\psi\in\Z\ .}
The quantization of $\onov{8\pi^2}tr(f\wedge f)$ depends on the choice of background, and the CS theory is not well defined by itself, but if as part of the $U(1)_\psi$ gauging, we add \eqref{etaA}, the full theory is periodic and well defined. In this way, the $\Z_N$ anomaly teaches us about the existence of a cusp and constrains the effective theory living on it even though $\Z_N$ is not a symmetry of our theory.

 What happens at finite $N$? We define $Y=\onov{\vev{\overline{\psi}_D\psi_D}}\overline{\psi}_D\psi_D$ as in \eqref{Y} and consider its potential $V(Y)$. The potential has a minimum at $Y=1$. Consider the setup in which we put the theory on $\mathcal{M}_4\times I$ with $\Z_N$ twisted boundary conditions on the two sides of the interval. Equivalently, we can turn on a small mass for $\psi_D$. This mass breaks the $\Z_N$ symmetry. By changing the phase of the mass continuously along the interval $I$, we can force the theory to be on different $\Z_N$ vacua on the two sides of the interval. We can connect the two vacua via the following trajectory. Our starting point is the vacuum $U_\la=1\ , Y=1$. We start changing the phase of $Y$ continuously until at some point ($arg(Y)=\pi/N$), we jump to $U_\la= e^{2\pi i/N}$ with $arg(Y)=-\pi/N$. Now we continue changing the phase of $Y$ until we reach the second vacuum with $Y=1$. The total phase acquired along this trajectory must be $\delta S_{tot}$ as in \eqref{deltaSs}. We know that the phase coming from jumping between the $U_\la$ vacua is $\delta S_\la$ and this tells us that the phase acquired by changing $Y$ must be $\delta S_\psi$. Repeating the argument of section \ref{sec:reviewQCD}, the theory must be single valued under $Y\to e^{2\pi i}Y$. If we can close a smooth loop along which $\int d\arg(Y)=2\pi$, then any coupling of the coordinate on the loop to background gauge fields can be removed by adding a local counterterm. Then, the anomaly matching condition \eqref{anomalycondition} cannot be satisfied. The most plausible resolution is that there is a cusp at $\arg(Y)=\pi$ as in $N_f\geq 2$ and in large $N$.

To summarize, the anomaly associated with the $\Z_N$ discrete axial symmetry of the bigger theory $\mathcal{T}_{x-ray}$ can be used to constrain the IR phase of $N_f=1$ QCD. The constraints are satisfied simply by taking $N_f=1$ QCD to behave the same as $N_f\geq 2$ QCD, even though there is no anomaly in $N_f=1$ QCD forcing it to exhibit this behaviour. The condensation of $\overline{\psi}_D\psi_D$ and the cusp on its potential from this perspective are now an essential ingredient in the confining phase of $N_f=1$ QCD, as the confining phase is inconsistent without it.  
\subsection{$N_f=1$ baryons and $\Z_N^2-U(1)_\psi$ anomaly}
\label{subsec:Nf1baryons}
\begin{figure}
	\vspace{4pt}
	\begin{center}
		\includegraphics[width=0.65\textwidth]{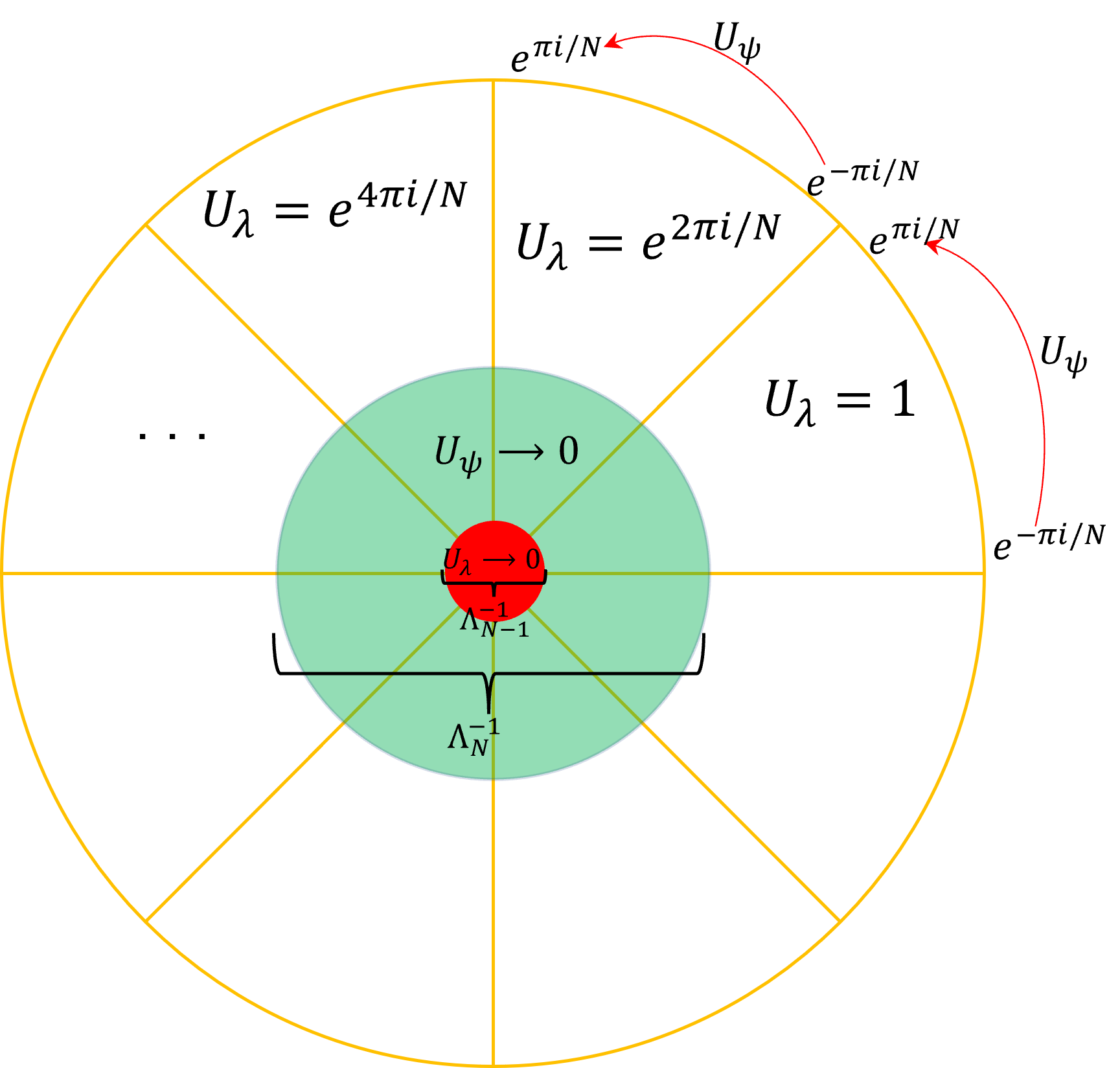}
		\caption{\footnotesize{An illustration of the junction configuration. The straight yellow lines represent domain walls in which the values of the two condensates jump $U_\la\to e^{2\pi i/N}U_\la\ ,\ U_\psi\to e^{-2\pi i/N}U_\psi$. Each "pizza slice" has a definite value for $U_\la$, while $U_\psi$ changes along the slice from $e^{-\pi i/N}\to e^{i\pi/N}$. $\Z_N$ axial transformations accompanied by space rotations leave the configuration invariant. On the junction, the condensates should go to zero and new massless modes are expected to appear. The inner green circle has  radius $\sim\Lambda_N^{-1}$. In this regime, $U_\psi\to0$ but $U_\la$ is still the same as in the outer regime. The inner red circle has radius $\sim\Lambda_{N-1}^{-1}\ll\Lambda_N^{-1}$ and this is where $U_\la\to 0$. }}\label{pizza}
	\end{center}
	\vspace{4pt}
\end{figure}

So far we discussed the implications of anomalies linear in $\Z_N$, such as $\Z_N\times U(1)_\psi^2$. However, there is also an anomaly quadratic in $\Z_N$ that should be matched by the $U_\psi$ condensate. It is given by the 5d Chern-Simons term \footnote{See for examle \cite{Kapustin:2014zva} for an explanation about CS terms involving discrete gauge fields.} \eql{zN2anomaly}{S_{Z_N^2-U(1)_\psi}=\onov{8\pi^2}\int_{\mathcal{M}_5}A_N\wedge dA_N\wedge dA_\psi\ ,}
where $A_N$ is the $\Z_N$ gauge field. See \ref{sec:detanomaly} for more details.
How is this realized in the IR when $\Z_N$ is broken? Anomalies linear in $\Z_N$ can be sometimes realized on domain walls from the following reason. These anomalies are described by a 5d CS term $\sim A_N\wedge Q_4$ where $Q_4$ is some topological density of the unbroken symmetry. Turning on non-trivial value $\int A_N=\frac{2\pi}{N}$ forces the vacuum to be in a domain wall configuration. We can integrate over the direction orthogonal to the domain wall. This results in  having a domain theory with the anomaly $\frac{2\pi}{N}Q_4$. We can try and repeat the same for $A_N\wedge dA_N\wedge dA_\psi$ and expect to find a $dA_N\wedge dA_\psi$ anomaly on the domain wall. However, this doesn't work since $\Z_N$ is broken and is not a symmetry of the domain wall theory. Instead of a domain wall configuration, the configuration we should study is the junction of domain walls, see figure \ref{pizza}. A similar configuration for Super Yang-Mills was studied in \cite{Gaiotto:2013gwa}.
 The advantage of this configuration is that the $\Z_N$ symmetry is restored on the effective 2d theory on the junction between the domain walls. In analogy with a domain wall, we can start from the 5d CS action $\onov{8\pi^2}\int_{\mathcal{M}_5}A_N\wedge dA_N\wedge dA_\psi$ that describes a 4d anomaly. We can reduce it to a 3d CS action that describes the 2d anomaly on the junction by integrating over the two orthogonal directions for which $\int dA_N=2\pi$ (we also need to multiply by an extra combinatoric factor of $2$). We find that the 2d theory on the junction should have $\Z_N\times U(1)$ symmetries with a mixed anomaly between them given by the 3d CS term $\onov{2\pi}A_N\wedge dA_\psi$. As in the domain wall case, also here the total $\Z_N^2\times U(1)_\psi$ anomaly of $\mathcal{T}_{x-ray}$ is zero. However, due to the scale separation $\Lambda_{N}\ll\Lambda_{N-1}$ we can split the anomaly to \eq{ S_\psi^{3d}=-S_\la^{3d}=\onov{2\pi}\int A_N\wedge dA_\psi\ .} $ S_\la^{3d}$ is expected to be matched by a 2d zero mode living on the red circle in \ref{pizza}. This can be shown by constructing the junction in the uv where $SU(N)$ is weakly coupled. Therefore, there must be a 2d zero mode living the green circle in \ref{pizza} to match the $S_\psi^{3d}$ anomaly. There are many ways to saturate such an anomaly in 2d, and we leave the study of this theory to future work. We will only comment on the relation of this anomaly to $N_f=1$ baryons. Interestingly, this anomaly has a similar structure to the $U(1)_B-SU(N_f)_L^2$ anomaly in $N_f\geq 2$ QCD discussed in \ref{subsec:skyrme}. It tells us that the zero mode on the $U_\psi\to 0$ singularity carries $U(1)_\psi$ charge. Therefore, a configuration with $U(1)_\psi$ singularity has excitations that correspond to $N_f=1$ baryons.
 How does such a configuration look like? Along one direction, $\eta'$ is forced to wind. Its winding leads to a singularity on which we find massless modes as explained above. Excitation of the massless mode along the singularity has baryon charge. This  is in agreement with the construction of $N_f=1$ baryons described in \cite{Komargodski:2018odf, Karasik:2020pwu}. Thus, our anomaly considerations give another argument in favour of the identification of $N_f=1$ baryons with the pancake configuration described in \cite{Komargodski:2018odf}. 

\subsection{Detailed analysis of the anomalies}
\label{sec:detanomaly}
In this section we give a detailed analysis of the anomalies mentioned through out section \ref{sec:Nf1qcd}.
The gauge+global symmetry group of $\mathcal{T}_{x-ray}$  is\footnote{In principle, we should also include spacetime symmetry $spin(4)$. However, there are no anomalies involving gravity in this case, simply because there are $N\Z$ fermions transforming with a phase of $2\pi/N$.} \eq{\frac{[SU(N-1)]\times U(1)_\la}{\Z_{N-1}}\times \frac{[SU(N)]\times U(1)_{\psi}}{\Z_N}\times (\Z_N)_L\ .}
We start by defining the dynamical fields $a_1\in SU(N)\ ,\ a_2\in SU(N-1)$, and their field strengths $f_1$ and $f_2$. Next we promote them to $U(N)$ and $U(N-1)$ gauge fields by writing \eq{\tilde{a}_1=a_1+\onov{N}A_\psi\ ,\ \tilde{a}_2=a_2+\onov{N-1}A_\la\ .}
The last step is to add the $(\Z_N)_L$ gauge field. This is done by introducing a pair of a U(1) gauge field $A_N$ and a periodic scalar $\phi$ satisfying the constraint $NA_N=d\phi$. The covariant derivatives of the different fermions are
\eq{&D\psi_L=(d-i\tilde{a}_1-iA_N)\psi_L\ ,\ D\psi_R=(d-i\tilde{a}_1)\psi_R\ ,\ \\&D\la_L=(d-i\tilde{a}_1-iA_N)\la_L+i\la_L\tilde{a}_2\ ,\ D\la_R=(d-i\tilde{a}_1-iA_N)\la_R+i\la_R\tilde{a}_2\ .}
We will also write explicitly the quantization conditions of the following integrals
\eq{&\onov{8\pi^2}\int_{\mathcal{M}_4}tr(\tilde{f}_{1,2}^2)\in\Z\ ,\ \onov{2\pi}\int_{\mathcal{M}_2}tr(\tilde{f}_{1,2})=\onov{2\pi}\int_{\mathcal{M}_2}dA_{\psi,\la}\in\Z\ ,\\& \onov{2\pi}\oint A_N=\onov{N}\Z\ ,\ \onov{2\pi}\int_{\mathcal{M}_2}dA_N\in\Z\ \text{with}\ NdA_N=0\ .}

The part of the anomaly carried by $\psi_D$ and $\la_D$ can be computed from their 6d anomaly polynomials
\eq{\mathcal{A}_\psi&=\onov{24\pi^2}tr_\psi[(\tilde{f}_1+dA_N)^3-(\tilde{f}_1)^3]\\
&=\onov{8\pi^2}dA_Ntr(\tilde{f}_1^2)+\onov{8\pi^2}dA_NdA_NdA_\psi+\frac{N}{24\pi^2}dA_NdA_NdA_N\ ,\\
\mathcal{A}_\la&=\onov{24\pi^2}tr_\la[(\tilde{f}_1+\tilde{f}_2+dA_N)^3-(\tilde{f}_1+\tilde{f}_2)^3]\\
&=\frac{N-1}{8\pi^2}dA_Ntr(\tilde{f}_1^2)+\frac{N}{8\pi^2}dA_Ntr(\tilde{f}_2^2)+\frac{1}{4\pi^2}dA_NdA_\psi dA_\la\\&+\frac{N-1}{8\pi^2}dA_NdA_NdA_\psi+\frac{N}{8\pi^2}dA_NdA_NdA_\la+\frac{N(N-1)}{24\pi^2}dA_NdA_NdA_N\ ,\\
\mathcal{A}_{tot}=\mathcal{A}_\psi+\mathcal{A}_\la&=\frac{N}{8\pi^2}dA_Ntr(\tilde{f}_1^2)+\frac{N}{8\pi^2}dA_Ntr(\tilde{f}_2^2)+\frac{1}{4\pi^2}dA_NdA_\psi dA_\la\\&+\frac{N}{8\pi^2}dA_NdA_NdA_\psi+\frac{N}{8\pi^2}dA_NdA_NdA_\la+\frac{N^2}{24\pi^2}dA_NdA_NdA_N\ .}
We can ask how does the partition function varies under a $\Z_N$ gauge transformation. This is equivalent to reducing the 6d anomaly polynomial to a 4d action by replacing one of the $dA_N$ by $\frac{2\pi}{N}$. We get
\eql{deltaSs}{\delta S_\psi&=\frac{2\pi}{8\pi^2N}\int tr(\tilde{f}_1)^2+\frac{2\pi}{4\pi^2N}\int dA_NdA_\psi\ mod\ 2\pi\ ,\\
\delta S_\la&=-\frac{2\pi}{8\pi^2 N}\int tr(\tilde{f}_1^2)+\frac{2\pi}{4\pi^2 N}\int dA_\psi dA_\la-\frac{2\pi}{4\pi^2N}\int dA_NdA_\psi\ mod\ 2\pi\ ,\\
\delta S_{tot}&=\frac{2\pi}{4\pi^2 N}\int dA_\psi dA_\la\ mod\ 2\pi\ .}
The only anomaly is the triangle $(\Z_N)_L\times U(1)_\psi\times U(1)_\la$ anomaly. Assuming $U(1)_{\psi,\la}$ are not broken, we expect $(\Z_N)_L$ to be broken by the vacuum, which is indeed what we find. At energies $E\ll\Lambda_2$, the effective theory contains the $U_\la$ condensate and the fermion $\psi_D$, weakly coupled to the $SU(N)$ gauge fields. $\psi_D$ still contributes $\delta S_\psi$ to the anomaly, which implies that the condensate $U_\la$ must contribute $\delta S_\la$. Let's see exactly how it works. In this limit, we can think of $U_\la$ as a condensate in $SU(N-1)$ QCD with $N$ flavours. A Domain wall connecting $U_\la=1$ and $U_\la=e^{2\pi i/N}$ is described by the effective 3d theory of a $SU(N-1)_{1-N/2}$ CS theory coupled to $N$ charged fermions \cite{Gaiotto:2017tne}. In our case, the $SU(N)$ symmetry acting on the fermions is gauged. The $SU(N)$ gauge fields appear on the domain wall with the fractional bare CS level $\frac{N-1}{N}$ as can be read off from the WZ action, see section \ref{sec:reviewQCD}. The fact that the level is fractional is not a problem in this case because there is another fractional contribution coming from the variation of $\psi_D$ which together sums to an integer. The theory has a global $U(1)$ symmetry that has a common $\Z_N\times \Z_{N-1}$ with the $SU(N)\times SU(N-1)$ gauge groups. To perform the quotients we introduce 2 pairs of $U(1)$ 1-form and 2-form gauge fields, satisfying the constraints
\eq{NB_N^{(2)}=dB^{(1)}_N\ ,\ (N-1)B_{N-1}^{(2)}=dB_{N-1}^{(1)}\ .}
They are invariant under the 1-form gauge redundancies
\eq{\delta B_N^{(1)}=d\omega_N\ ,\ \delta B_{N-1}^{(1)}=d\omega_{N-1}\ ,\ \delta B_N^{(2)}=N\omega_N\ ,\ \delta B_{N-1}^{(2)}=(N-1)\omega_{N-1}\ .} 
Next, we promote the $SU(N),\ SU(N-1)$  gauge fields to $U(N)
,\ U(N-1)$ gauge fields by defining
\eq{\tilde{a}_1=a_1+\onov{N}B_N^{(1)}\ ,\ \tilde{a}_2=a_2+\onov{N-1}B_{N-1}^{(1)}\ .} 
Similarly, we define the background $U(1)$ gauge field to be 
\eq{\tilde{A}_V=A_V-\onov{N}B_N^{(1)}-\onov{N-1}B_{N-1}^{(1)}\ .}
 
The CS part of the theory in the presence of the background gauge fields is
\eq{&\frac{N-1}{4\pi N}tr(\tilde{a}_1d\tilde{a}_1-2i/3\tilde{a}_1^3)+\frac{1}{4\pi}tr(\tilde{a}_2d\tilde{a}_2-2i/3\tilde{a}_2^3)+\frac{p}{4\pi}A_VdA_V-\frac{2p}{4\pi N}A_VdB_N^{(1)}-\frac{2p}{4\pi (N-1)}A_VdB_{N-1}^{(1)}\\&+\frac{2p}{4\pi N(N-1)}B_N^{(1)}dB_{N-1}^{(1)}+\frac{p-N+1}{4\pi N^2}B_N^{(1)}dB_N^{(1)}+\frac{p-N+1}{4\pi(N-1)^2}B_{N-1}^{(1)}dB_{N-1}^{(1)}\ ,}
where we used the bare levels which give the correct quantization conditions, and added a counterterm for the $U(1)$ gauge field with arbitrary coefficient $p$. The role of the counterterm in this procedure is the minimize the order of the anomaly. Since there are $N$ vacua, we know that the order of the anomaly cannot exceed $N$. indeed, by choosing $p=N-1$, we get 
\eq{&\frac{N-1}{4\pi N}tr(\tilde{a}_1d\tilde{a}_1-2i/3\tilde{a}_1^3)+\frac{1}{4\pi}tr(\tilde{a}_2d\tilde{a}_2-2i/3\tilde{a}_2^3)+\frac{N-1}{4\pi}A_VdA_V-\frac{2(N-1)}{4\pi N}A_VdB_N^{(1)}\\&-\frac{2}{4\pi }A_VdB_{N-1}^{(1)}+\frac{2}{4\pi N}B_N^{(1)}dB_{N-1}^{(1)}\ .}
The fractional part matches most of the anomalous contribution that is expected to be carried by $U_\la$ as can be read off from $\delta S_\la$ in \eqref{deltaSs}. The only missing part is the last term $-\frac{2\pi}{4\pi^2N}\int dA_N dA_\psi$. This term comes from an anomaly quadratic in $(\Z_N)_L$ and cannot be observed on the domain wall. The reason is that $(\Z_N)_L$ is not a symmetry of the domain wall. To observe anomalies quadratic in $(\Z_N)_L$ we need to study a junction of $N$ domain walls, see appendix \ref{sec:junctionanomalies} for more details. On the 2d junction theory, the $(\Z_N)_L$ symmetry is restored and we expect to find the anomaly described by the 4d anomaly polynomial $-\frac{2\pi}{4\pi^2}\int dA_N dA_\psi$. This should work exactly as described in section \ref{subsec:Nf1baryons}.

\section{Chiral anti-symmetric theories}
\label{sec:AS}
In this section we will use the same procedure advocated in the previous sections, to derive constraints on certain chiral gauge theories. In analogy with $\overline{\psi}_D\psi_D$ in $N_f=1$ QCD, we can define a "chiral condensate" operator we denote by $\mathcal{B}$ which is charged under discrete $\Z_N$ axial transformations of $\mathcal{T}_{x-ray}$. As we will show, this condensate has applications to large $N$ dynamics \ref{subsec:basicAS} and to predicting phases of certain theories \ref{subsec:multiAS}.
\subsection{The basic antisymmetric theory}
\label{subsec:basicAS}
In this section we will apply our method on one of the simplest chiral gauge theories. Consider an $SU(N)$ gauge theory coupled to one left-Weyl fermion\footnote{Throughout the paper, fermions without a $D$ subscript are left-handed Weyl fermions, while with a $D$ subscript are Dirac fermions.} $\chi$ in the $\anti$ representation, and $N-4$ left-Weyl fermions $\psi$ in the $\overline{\Box}$ representation. This theory has a global symmetry of $SU(N-4)_\psi\times U(1)$. $SU(N-4)_\psi$ acts on the $N-4$ $\psi$s, and the $U(1)$ charges of the fields are
\eq{q_\chi=N-4\ ,\ q_\psi=2-N\ .} One of the interesting properties of this theory is that all the anomalies can be matched by the massless composite gauge invariant fermion $\chi\psi\psi$ \cite{Dimopoulos:1980hn}. This fermion is in the $\sym$ representation of $SU(N-4)_\psi$ and has $U(1)$ charge of $q_{\chi\psi\psi}=-N$. Hence, a phase without any symmetry breaking is consistent and serves as a plausible candidate for the IR theory. Moreover, it has been argued in \cite{Eichten:1985fs} that assuming confinement, the global $U(1)$ cannot be broken in the large $N$ limit. Then the triangle and gravitational anomalies associated with the $U(1)$ must be matched by massless gauge invariant fermions. The fermions needed to match the $U(1)$ anomalies are $\chi\psi\psi$ which automatically match all the other anomalies of the theory. This gives another strong argument in favour of the phase without symmetry breaking \footnote{The authors of \cite{Bolognesi:2020mpe,Bolognesi:2021yni} argued that there is a new anomaly forbidding this phase, however in \cite{Smith:2021vbf} it was shown that this anomaly is fictitious and the phase without symmetry breaking is consistent.}.

We can deform this theory in the uv by adding an $SU(N-1)$ gauge theory and a bifundamental Dirac fermion $\la_D$. The full theory enjoys a discrete $\Z_N$ axial symmetry that acts as
\eql{basicaxial}{\la_D\to e^{i\pi \gamma_5/N}\la_D\ ,\ \psi\to e^{2\pi i/N^2}\psi\ ,\ \chi\to e^{-4\pi i/N^2}\chi\ .}
This transformation generates $\Z_N$ in the sense that acting with it $N$ times, gives \eq{\la_D\to-\la_D\ ,\ \psi\to e^{2\pi i/N}\psi\ ,\ \chi\to e^{-4\pi i/N}\chi\ ,}
which can be undone by the actions of $U(1)_\la$ and the centre of the $SU(N)$ gauge group. 
  This axial symmetry has several anomalies. For example, under this axial transformation, the theta angle for background $SU(N-4)$ gauge fields is shifted by $\delta \te_{N-4}=\frac{2\pi}{N}$. Again, we will take the limit in which the $SU(N-1)$ confines when the $SU(N)$ is still weakly coupled. The intermediate theory will be $SU(N)+\chi+\psi$ and in addition, an $SU(N)$ NLSM coming from the $\overline{\la}_D\la_D$ condensate, parametrized by $U_\la\in SU(N)$. The target space is lifted except for the central elements $U_\la=e^{2\pi ik/N}\boldmath{1}$. The $\te$ angle in the $k$ vacuum is $\te_k=\frac{2\pi k}{N}$. This theta angle can be rotated away using axial rotations. Assume that the vacuum of the chiral theory contains the massless fermions $\chi\psi\psi$ without breaking any symmetries. The massless composite fermions are not charged under the axial rotations \eqref{basicaxial} and therefore cannot carry the anomalies associated with it. As in $N_f=1$ QCD, consistency of this phase implies that there exists a condensate that doesn't break any global symmetry but does break $\Z_N$.
There is a unique such operator, $\mathcal{B}\sim \chi^{N-2}\psi^{N-4}$. First let us specify how all the indices are contracted. There are $2N-6$ spinor indices which are contracted to give a scalar, $N-4$ indices of $SU(N-4)$ which are contracted using the epsilon tensor. Finally, there are $2N-4$ fundamental indices and $N-4$ antifundamental indices of $SU(N)$. $N-4$ are paired and the residual $N$ are contracted using the epsilon tensor. The $U(1)$ charge of this operator is $(N-2)(N-4)+(N-4)(2-N)=0$. Therefore it is neutral under all the global symmetries. Under the discrete axial rotation \eqref{basicaxial} it transforms as
\eq{\mathcal{B}\to e^{-\frac{2\pi i}{N}}\mathcal{B}\ .}
Similar to the arguments used in section \ref{sec:Nf1qcd}, the mixed anomaly between the discrete axial symmetry and $SU(N-4)$ (as an example) implies that in the phase without symmetry breaking, $\mathcal{B}$ must condense. The phase of $\mathcal{B}$ then matches the anomaly, similar to the phase of $\overline{\psi}_D\psi_D$ in QCD. Without its condensation, nothing in the IR can match the anomaly and the phase is inconsistent. 
Why is the operator $\mathcal{B}$ interesting? One may wonder whether the condensation of $\mathcal{B}$ can occur in the large $N$ limit. According to \cite{Eichten:1985fs}, there is no operator charged under the global $U(1)$ which can condense in the large $N$ limit. The reason is that such operators cannot appear in the cut of a planar diagram. The common lore as stated in \cite{Eichten:1985fs} is that operators whose color indices are contracted using an $\epsilon$ tensor cannot condense in the large $N$ limit. This includes also $\mathcal{B}$. Therefore, if the large $N$ analysis is correct, the $U(1)$ cannot be broken and also $\mathcal{B}$ cannot condense. This leaves us with no consistent confining phase in the large $N$ limit. There are two possibilities. The first is that the theory is not confining \footnote{Recently, it was claimed in \cite{Anber:2021iip} that operators made out of large number of fermions will most likely not condense, and used it as an argument supporting deconfinement in certain theories. However, we don't see any general problem in having such condensates when flowing to strong interactions.   }. While we cannot disprove it, this seems unlikely since we are far from the asymptotic freedom bound, see equation \eqref{beta}. The second possibility is that the large $N$ analysis of \cite{Eichten:1985fs} is incorrect. We want to propose that indeed there is a loophole in the large $N$ analysis of \cite{Eichten:1985fs} and that there is nothing wrong with the condensation of $\mathcal{B}$ and various $U(1)$ charged operators. Basically, this is because the number of fermions scales like $N^2$ in this theories which allows for diagrams with large number of internal fermions. Using identities that relate the $\epsilon$ tensor to product of $\delta$s, it is possible to show that these operators can appear in the cuts of leading diagrams. The details and a refined large $N$ analysis is left for future work \cite{KOT}. 
	
\subsection{x-ray uv mixing: A lesson from a Higgs phase}
\label{subsec:mixing}
When analysing the constraints coming from the $\Z_N$ symmetry, we assumed that we have a scale separation between the "x-ray" and the "uv" degrees of freedom. In the uv, we can compute independently the contribution to the anomaly coming from the $\bar{\la}_D\la_D$ condensate and from our uv fields $\psi,\chi$. Assuming $\bar{\la}_D\la_D$ stays untouched, we can forget about it and study $\mathcal{T}_{uv}$ by itself. However, the situation is more subtle. Consider a scalar operator $\mathcal{O}$ in the fundamental of the $SU(N)$ gauge group made out of uv fields. This operator can dress the vacuum by replacing the condensate \eql{dressing}{\bar{\la}_D\la_D\to \mathcal{O}\bar{\la}_D\la_D\mathcal{O}^\dagger\ .}
If this new vacuum can match all of the constraints, the condensate of $\mathcal{B}$ is not needed. We will show first that such a dressing can happen, and then we will argue that while it carries some of the anomalies and changes the constraints on the IR, it cannot capture all of them so the $\mathcal{B}$ condensate is still needed.
It is well known that the IR phase of the chiral theory can be obtained via a Higgs phase description using the idea of complementarity \cite{Dimopoulos:1980hn}. The logic behind complementarity is that two phases are equivalent and indistinguishable if they share the same massless physical spectrum even if the description may seem very different. For example, one phase involves strong interactions and confinement while the other phase involves weak interactions and Higgsing. In addition to the massless spectrum, the two theories should have the same topological order or vacuum degeneracy in case it is not trivial. Vevs of gauge invariant operators that break some symmetries should be matched, but vevs of neutral operators are usually ignored. Needless to say that a gauge dependent vev of a Higgs field doesn't need to have any physical analogue on the dual side. However, we would like to claim that vevs of neutral operators that transform non-trivially under the axial $\Z_N$ are a necessary ingredient in the matching of the two phases \footnote{The importance of such condensates was also noticed in \cite{Bolognesi:2021hmg}. However, there it was used to argue incorrectly that the Higgs and the confining phases are different instead of identifying the dual condensate on the confining side.}. Even though they don't break any symmetry, they carry some anomalies of the bigger bifundamental theory $\mathcal{T}_{x-ray}$ which makes them part of the IR data. Moreover, we claim that it should be impossible to reach the studied IR phases without a condensate charged under the relevant axial transformation (or with a different object saturating the anomaly in case it is possible). In addition, vevs of gauge dependent Higgs fields also have a physical meaning. They are equivalent to dressing the vacuum as in \eqref{dressing} with the Higgs field. This can be seen from the color-flavour locking pattern imposed by the Higgs condensate. Its physical importance is by changing the constraints needed to be satisfied by the $\mathcal{B}$ condensate which affect the coupling of $arg(\mathcal{B})$ to background gauge fields.
 To show how it works consider adding to our chiral theory the scalar Higgs field $\phi$ with the Yukawa interaction $\phi\chi\psi+c.c$. Unlike the $N_f=1$ QCD case, here the scalar $\phi$ is charged under the gauge and global symmetries. Its representations under $[SU(N)]_{gauge}\times SU(N-4)\times U(1)$ are $(\overline{\Box},\overline{\Box})_2$. We can give $\phi$ a vev of the form $\phi_{ia}\sim\delta_{ia}$ where $i,a$ are flavour and colour indices respectively. This Higgses the gauge group down to $SU(N)\to SU(4)$ and locks the global symmetry with broken gauge transformations. Notice that also the discrete axial transformation $\phi\to e^{2\pi i/N^2}\phi$ leave the vacuum invariant when accompanied by the $SU(N)$ gauge transformation \eql{axgauge}{\phi\to \phi U^\dagger\ ,\ U=\mat{e^{\frac{2\pi i}{N^2}}\boldmath{1}_{N-4}&\\&e^{-\frac{\pi i(N-4)}{2N^2}}\boldmath{1}_4}\ .} 

The massless fermions that remain are:
\begin{itemize}
	\item One fermion in the $(1,\sym)_{-N}$ under $[SU(4)]_{gauge}\times SU(N-4)\times U(1)$. It comes from the fermion $\psi$ where its color index becomes a flavor index due to the color-flavor locking. In addition, this fermion is invariant under the $\Z_N$ axial transformation (accompanied by the gauge transformation \eqref{axgauge}). This fermion has exactly the same quantum numbers as the composite fermion $\chi\psi\psi$.
	\item One fermion in the $\left(\anti,1\right)_0$ under $[SU(4)]_{gauge}\times SU(N-4)\times U(1)$. It comes from the fermion $\chi$. Under the action of $\Z_N$ it transforms as \eql{su4chi}{\chi\to e^{-\frac{4\pi i}{N^2}}e^{-\frac{\pi i(N-4)}{N^2}}\chi=e^{-\frac{\pi i}{N}}\chi}
\end{itemize} 
As we continue flowing down, the $SU(4)$ gauge group becomes strongly coupled and is expected to confine. It is expected that the fermion bilinear $\chi\chi$ condenses. After the $SU(4)$ confinement we are left with the $\left(\anti,1\right)_0$ massless fermion and the condensate of $\chi\chi$. This condensate doesn't break any symmetry but transforms under the $\Z_N$ axial symmetry as $\chi\chi\to e^{-\frac{2\pi i}{N}}\chi\chi$. The gauge invariant description of $\chi\chi$ must be the operator $\mathcal{B}$, so we see that indeed $\vev{\mathcal{B}}\neq 0$. However, this raises a question. $\chi$ itself is not charged under the global $SU(N-4)\times U(1)$. This implies that $\chi$ cannot contribute to mixed $\Z_N-SU(N-4)$ and mixed $\Z_N-U(1)$ anomalies, and so does the condensate $\chi\chi$. If this is true, then what carries these anomalies? The answer is that the Higgs vacuum with $\vev{\chi\psi}\neq 0$ can be interpreted as replacing the $U_\la=\bar{\la}_D\la_D$ condensate by $U_{\la\chi\psi}=\psi\chi\bar{\la}_D\la\chi^\dagger\psi^\dagger$. If the new condensate $U_{\la\chi\psi}$ carries the anomaly, then why do we need the $\mathcal{B}$ condensate? The answer is that it doesn't carry all the anomalies. The easiest way to understand the anomalies left to be carried by $\mathcal{B}$ is to look at the anomalies associated with the transformation \eqref{su4chi}. This transformation has a $\Z_N^3$ triangle anomaly and a $\Z_N-gravity$ anomaly.  If we could have found an operator $\mathcal{O}$ such that $U_{\la\mathcal{O}}=\mathcal{O}\bar{\la}\la\mathcal{O}^\dagger$ carries all the anomalies of $\mathcal{T}_{x-ray}$, then indeed the condensate of $\mathcal{B}$ would be unnecessary. However, this is not the case here and we conclude that in the phase without symmetry breaking, $\mathcal{B}$ must condense. As explained in appendix \ref{sec:junctionanomalies}, anomalies of the type $\Z_N\times gravity $ and $\Z_N^3$ are realized on junctions of domain walls, and a cusp is not needed in this case.

\subsection{Multiflavor chiral theory}
\label{subsec:multiAS}
In this section we will study the multiflavor version of the basic chiral theory studied in \ref{subsec:basicAS}. We will use our knowledge about the $\mathcal{B}$ condensate of the basic theory to predict the IR phase in its multiflavor version.
Consider the $SU(N)$ gauge theory coupled to $K$ antisymmetric fermions $\chi$ and $K(N-4)$ anti-fundamentals fermions $\psi$. The 1-loop beta function is proportional to
\eql{beta}{\beta_N\sim -11N+K(N-4)+K(N-2)=-11N+2KN-6K\ .}
Asymptotic freedom is guaranteed when $2KN<6K+11N$. The fate of this theory in the IR is not known. Close to the asymptotic freedom bound, the theory might flow to a non-trivial conformal fixed point, but for small enough $K$ it is plausible that the theory confines. We will assume that there exist some value of $K>1$ for which the theory confines, and study the constraints on the IR.

In this theory there is a global symmetry of $SU(K)_{\chi}\times SU(K(N-4))_{\psi}\times U(1)$. Under the $U(1)$, the charges of the fields are $q_\chi=N-4\ ,\ q_\psi=2-N$. As specified in \ref{multianomalies}, there are many anomalies in this theory. How can all of them be matched in a confining phase? There is more than one way to do it, but a relatively simple way is to start from what we know about the $K=1$ case of section \ref{subsec:basicAS} and assume that also here $\mathcal{B}=\chi^{N-2}\psi^{N-4}$ condenses and that the fermions $\chi\psi\psi$ remain massless. Here we need to be more specific about the indices participating in these two objects.
\begin{itemize}
	\item $\mathcal{B}$: As in the $K=1$ case, it is gauge invariant, Lorentz scalar and $U(1)$ neutral. The difference is in its flavor indices. It has $N-2$ $SU(K)_\chi$ indices and $N-4$ $SU(K(N-4))_\psi$ indices. We can split the $SU(K(N-4))_\psi$ index into an index $I=1,...,K$ parametrizing $SU(K)_\psi$ and $a=1,...,N-4$ parametrizing $SU(N-4)_\psi$, such that every $\psi$ carries both $I,a$ as its flavor indices. The $N-4$ $SU(N-4)_\psi$ indices can be contracted using an epsilon tensor. The $N-4$ $SU(K)_\psi$ indices can be paired with $N-4$ of the $SU(K)_\chi$. This breaks $SU(K)_\chi\times SU(K)_\psi\to SU(K)_{diag}$. However, we are left with two $SU(K)_\chi$ indices. For $K=2$, we can simply pair these together without breaking any additional symmetry. For $K>2$, we can pair them together but this will additionally break $SU(K)_{diag}\to SO(K)_{diag}$. To summarize, the condensate breaks
		\eql{breaking}{&SU(K)_\chi\times SU(K(N-4))_\psi\times U(1)\to SO(K)_{diag}\times SU(N-4)_\psi\times U(1)\ ,\ K>2\\
			&SU(K)_\chi\times SU(K(N-4))_\psi\times U(1)\to SU(K)_{diag}\times SU(N-4)_\psi\times U(1)\ ,\ K=2\ .}
	\item $\chi\psi\psi$: This fermion carries one $SU(K)_\chi$ index and two $SU(K(N-4))_\psi$ indices. In terms of the unbroken vacuum symmetries, these are three $SO(K)_{diag}$ indices and two $SU(N-4)_\psi$ indices. Two $SO(K)_{diag}$ indices are contracted such that the fermion is in the $(\ydiagram{1},\sym)_{-N}$ under $SO(K)_{diag}\times SU(N-4)_\psi\times U(1)$. For $K=2$ it works the same with the replacement of $SO(K)_{diag}\to SU(K)_{diag}$.
\end{itemize}
 As shown in \ref{multianomalies}, this phase is consistent with all the anomalies of the theory. As mentioned above, the proposal of this phase is inspired by the condensate of $\mathcal{B}$ in the $K=1$ theory.  Without taking it into considerations, the phase of the $K\geq 2$ theory looks unrelated with totally different dynamics. Thanks to the $\Z_N$ anomaly which predicts the $\mathcal{B}$ condensate also in the $K=1$ case, the theory for every value of $K$ is controlled by the same condensate (assuming confinement). Again we can ask about the large $N$ limit. The arguments of \cite{Eichten:1985fs} imply that the $U(1)$ cannot be broken in the confining phase in the large $N$ limit. The same arguments exactly can forbid condensation of $\mathcal{B}$ also here. One possibility is that the assumption of confinement is wrong and the theory flows to a non-trivial fixed point already for $K=2$ (at least for large $N$). While this cannot be ruled out, $K=2$ is still pretty far from the asymptotic freedom bound. Using the new constraints and the analysis in section \ref{subsec:basicAS}, we were able to promote the contradiction to $K=1$ where a deconfining phase with a non-trivial fixed point is even more unlikely. We want to claim that there is no problem with the suggested phase in the large $N$ limit, because the argument made in \cite{Eichten:1985fs} is not correct, as was explained in \ref{subsec:basicAS}.
 
Notice that the suggested phase here can be also obtained by adding a Higgs field $\phi$ in the $(\overline{\Box},\overline{\Box},\overline{\Box})_2$ under $[SU(N)]_{gauge}\times SU(K)_\chi\times SU(K(N-4))_\psi\times U(1)$ with the Yukawa interaction $\phi\chi\psi+c.c.$. This is very similar to \ref{subsec:mixing}. We can split the $SU(K(N-4))_\psi$ index to $I=1,...,K$ and $a=1,...,N-4$. We give a vev to $\phi$ of the form $\vev{\phi}_{I,a,J,i}=\delta_{IJ}\delta_{a,i}$, where $i=1,...,N$ is the color index and $J=1,...,K$ is $SU(K)_\chi$ index. The vacuum preserves $SU(K)_{diag}\times SU(N-4)_\psi\times U(1)$ and an $SU(4)$ gauge group. The massless fermionic spectrum that remains contains $\psi'\sim\left(1,\overline{\ydiagram{1}},\sym\right)_{-N}$, and $\chi'\sim\left(\anti,\Box,1\right)_{0}$ under $[SU(4)]_{gauge}\times SU(K)_{diag}\times SU(N-4)_\psi\times U(1)$. Confinement of $SU(4)$ leads to a condensate of the charged fermions $\vev{\chi'\chi'}\neq0$, which results in a spontaneous breaking $SU(K)_{diag}\to SO(K)$ if $K>2$. 
Notice that before the confinement of $SU(4)$ and the breaking of $SU(K)_{diag}\to SO(K)$, we had 6 extra fermions in the fundamental of $SU(K)_{diag}$. These are exactly the missing 6 fermions needed to saturate the $SU(K)_{diag}^3$ anomaly which make the phase consistent also at intermediate regime before the confinement of $SU(4)$ as explained in appendix \ref{multianomalies}.

\section{The chiral symmetric theory}
\label{sec:sym}
The analysis made in section \ref{sec:AS} can be repeated for  similar chiral theories with a $\sym$ instead of $\anti$ fermion. All the details are the same up to some relative minus signs. Therefore, we will keep this section very concise and just go briefly over the symmetries and expected phases and condensates. All the conclusions and comments that were mentioned in \ref{sec:AS} apply also to this section. 
We will start from the basic chiral symmetric theory, which is the analogue of \ref{subsec:basicAS}. This theory contains an $SU(N)$ gauge group coupled to a fermion $\chi$ in the $\sym$ and $N+4$ fermions $\psi$ in the $\overline{\Box}$. The $\psi$s are charged under a global $SU(N+4)_\psi$. In addition, there is a global $U(1)$ with the charges 
\eq{q_\chi=N+4\ ,\ q_\psi=-N-2\ .} 
Similar to \ref{subsec:basicAS}, also this theory has a consistent phase without symmetry breaking. In this phase all the anomalies are matched by the composite fermions $\chi\psi\psi$ which sit in the $\anti$ representation under $SU(N+4)_\psi$ and have charge $q_{\chi\psi\psi}=-N$ under the $U(1)$. As done through out the paper, by adding an $SU(N-1)$ gauge group and a bifundamental Dirac field, we can conclude that in this phase there must be a condensate charged under the $\Z_N$ transformation,
\eq{\psi\to e^{2\pi i/N^2}\psi\ ,\ \chi\to e^{-4\pi i/N^2}\chi\ .}
There is a unique such operator invariant under all the symmetries of the theory, which is
\eq{\mathcal{B}=\chi^{N+2}\psi^{N+4}\ ,\ \Z_N:\ \mathcal{B}\to e^{-2\pi i/N}\mathcal{B}\ .}
Next, we will move on to the multiflavor symmetric theory. Here we take $K$ copies of the fermion content of the basic symmetric theory. We propose the phase in which $\mathcal{B}$ condenses. This leads to the symmetry breaking pattern
\eq{&SU(K)_\chi\times SU(K(N+4))_\psi\times U(1)\to SO(K)_{diag}\times SU(N+4)_\psi\times U(1)\ ,\ K>2\\&SU(K)_\chi\times SU(K(N+4))_\psi\times U(1)\to SU(K)_{diag}\times SU(N+4)_\psi\times U(1)\ ,\ K=2 \ .}
In addition, the composite massless fermions $\chi\psi\psi$ in the $\left(\Box,\anti\right)_{-N}$ under the unbroken symmetry saturates the rest of the anomalies. 

\section{Two gauge groups and phase transitions}
\label{sec:2gaugegroups}
In this section we will analyse several theories with two gauge groups. Such theories depend on a continuous parameter- the ratio of the two strong couplings $\Lambda_1/\Lambda_2$. Usually such theories can be analysed in the two limits $\Lambda_1/\Lambda_2\to 0,\infty$ while the intermediate regime remains mysterious. Using insights from our new $\Z_N$ anomaly, we give new proposals for the IR phase of several theories in sections \ref{sec:Georgi} and \ref{subsec:LT}. 
\subsection{Multiflavour Georgi model}
\label{sec:Georgi}
Our next examples are the Georgi model \cite{Georgi:1985hf} and its multiflavour generalization. Interestingly, these theories exhibit a similar behaviour as the chiral theories of \ref{sec:AS}, \ref{sec:sym} (and also as QCD). By this we mean that the basic theory has a consistent phase without symmetry breaking. However, the constraints coming from the $\Z_N$ anomaly imply that a certain operator $\mathcal{B}$ must condense. In the multiflavour case, it is the same operator $\mathcal{B}$ whose vev breaks the symmetry in a way that can give a consistent phase in the IR. We will start from the basic Georgi model. This is an $SU(N)\times SU(M)$ gauge theory with the left handed fermions,
\eq{\psi\ \text{in}\ (\overline{\Box},\Box)\ ,\ M\ \text{copies of}\ \xi\ \text{in}\ (\Box,1)\ ,\ N\ \text{copies of}\ \eta\ \text{in}\ (1,\overline{\Box})\ .}
This theory enjoys a global symmetry of $SU(N)_\eta\times SU(M)_\xi\times U(1)$, where under the $U(1)$,
\eq{q_\eta=q_\xi=-q_\psi=1\ .}
This theory also has the property that all the anomalies can be matched without symmetry breaking via the gauge invariant massless fermions, $\eta\psi\xi$ in the $(\Box,\Box)_1$ under $SU(N)_\eta\times SU(M)_\xi\times U(1)$. This theory can be analysed easily in the limits where one gauge group confines when the second is still weakly coupled. In one limit $\Lambda_N\gg\Lambda_M$, the $SU(N)$ confines when the $SU(M)$ is weakly coupled. The effective theory at intermediate energies $\Lambda_M\ll E\ll\Lambda_N$ is an $SU(M)$ gauge theory coupled to $\eta$ and to the $SU(M)$ NLSM $U\sim \xi\psi$. The vev of $U$ Higgses the $SU(M)$ completely and locks the global $SU(M)_{\xi}$ with $SU(M)$ gauge transformations. At low energies we are left with the fermion $\eta$ in the $(\Box,\Box)_1$ under $SU(N)_\eta\times SU(M)_\xi\times U(1)$ as $\eta\psi\xi$. Similarly, in the opposite limit we get in the IR the massless fermion $\xi$ in the same representations $(\Box,\Box)_1$. Their quantum numbers show that the gauge invariant description of the massless $\xi$ and $\eta$ in these two phases is $\eta\psi\xi$. The fact that the two limits give the same phase makes it even more plausible that this is the correct phase for every value of $\Lambda_N/\Lambda_M$.  Now we will move on to the multiflavor version of this theory by simply taking $K$ copies of every field. The global symmetry now is $SU(KN)_\eta\times SU(KM)_\xi\times SU(K)_\psi\times U(1)$. In this theory a phase without symmetry breaking cannot be found. We can start by analysing the two limits. The generalization is straight forward. In the $\Lambda_N\gg\Lambda_M$ limit the vev of $\xi\psi$ Higgses the $SU(M)$ as before and breaks \eql{Nbreak}{SU(KN)_\eta\times SU(KM)_\xi\times SU(K)_\psi\times U(1)\to SU(KN)_\eta\times SU(M)_\xi\times SU(K)_{\xi\psi}\times U(1)\ .} We are left in the IR with $\eta$ in the $(\Box, \Box,1)_1$ under the preserved symmetry. Similarly, in the $\Lambda_M\gg\Lambda_N$ limit, the vev of $\eta\psi$ breaks the symmetry as
\eql{Mbreak}{SU(KN)_\eta\times SU(KM)_\xi\times SU(K)_\psi\times U(1)\to SU(N)_\eta\times SU(KM)_\xi\times SU(K)_{\eta\psi}\times U(1)\ ,}
with the massless fermion $\xi$ in the $(\Box,\Box,1)_1$ under the preserved symmetry. Here the situation is different. The two limits give different symmetry breaking patterns, which means that there must be some kind of phase transition for intermediate $\Lambda_N/\Lambda_M$. The order and the properties of this phase transition are not known and are very hard to tackle. We will use our new anomaly conditions to propose a scenario for the phase transition. 
We can go back to the $K=1$ case. Since there are two gauge groups, there are two possible "x-ray" completions which lead to two independent constraints on the IR. We can define the following discrete $\Z_{N,M}$ axial transformations,
\eq{&\Z_N:\ \eta\to e^{-\frac{2\pi i}{MN}}\eta\ ,\ \psi\to e^{\frac{2\pi i}{MN}}\psi\ ,\ \xi\to\xi\ ,\ \delta\te_N=\frac{2\pi}{N}\ ,\ \delta\te_M=0\ ,\\&\Z_M:\ \xi\to e^{-\frac{2\pi i}{MN}}\xi\ ,\ \psi\to e^{\frac{2\pi i}{MN}}\psi\ ,\ \eta\to\eta\ ,\ \delta\te_M=\frac{2\pi}{M}\ ,\ \delta\te_N=0\ .}
Using the same techniques as in the previous sections we conclude that something in the IR must carry anomalies associated with them. There are several possible phases that we can construct. We will start from the phases in the two limits discussed above.

\begin{enumerate}
	\item $\Lambda_N\gg\Lambda_M$: In this limit we have the condensate $\vev{\psi\xi}\neq0$ and the massless fermion $\eta$. $\psi\xi$ transforms under the $\Z_N$ but invariant under the $\Z_M$. The gauge invariant description of this operator is therefore $\mathcal{B}_N=(\xi\psi)^M$. It looks like there is nothing that can carry anomalies involving $\Z_M$. However, the only anomaly that should be matched is $\Z_M\times SU(M)_\xi^2$ and this is matched by dressing the vacuum $\bar{\la}\la$ with the operator $\psi\xi$. See section \ref{subsec:mixing} for a detailed explanation.
	\item $\Lambda_M\gg\Lambda_N$: This is the same as the opposite limit but with the condensate of $\mathcal{B}_M=(\eta\psi)^N$ which is charged under $\Z_M$. The $\Z_N\times SU(N)_\eta^2$ anomaly is matched by dressing the vacuum $\bar{\la}\la$ with the operator $\psi\xi$.
	\item Intermediate $\Lambda_N/\Lambda_M$: There should be some kind of transition of the condensates in the two limits. We propose that there is a regime in the middle in which both $\mathcal{B}_M$ and $\mathcal{B}_N$ condense. 
	
\end{enumerate}

According to this proposal, the $K=1$ theory has different condensates as a function of the parameters, but without phase transition. Notice that without the $\Z_{N,M}$ symmetries of $\mathcal{T}_{x-ray}$, we don't give a lot of meaning to the distinction between the two condensates $\mathcal{B}_{N,M}$ as the two of them are neutral under all the symmetries.
Now we can move on to the $K>1$ case. We will conjecture that exactly the same operators condense as a function of $\Lambda_N/\Lambda_M$. The main difference is that now these operators break some of the global symmetry. This gives a natural way to connect the two limits also in the $K>1$ case. We always have the $NMK$ massless fermions $\eta\psi\xi$. In the $\Lambda_N\gg\Lambda_M$ limit, $\vev{\mathcal{B}_N}\neq 0,\ \vev{\mathcal{B}_M}=0$. This gives the symmetry breaking pattern \eqref{Nbreak} and the expected phase in this limit. Similarly, in the $\Lambda_M\gg\Lambda_N$ limit, $\vev{\mathcal{B}_M}\neq 0,\ \vev{\mathcal{B}_N}=0$ which gives \eqref{Mbreak} and the expected phase in this limit. 
For intermediate $\Lambda_N/\Lambda_M$ the two condensates are non-zero. This breaks the symmetry down to
\eq{SU(KN)_\eta\times SU(KM)_\xi\times SU(K)_\psi\times U(1)\to SU(N)_\eta\times SU(M)_\xi\times SU(K)_{diag}\times U(1)\ ,}
with $\eta\psi\xi$ in the $(\Box,\Box,\Box)_1$.
\subsection{Lohitsiri-Tong chiral models}
\label{subsec:LT}
In this section we will analyse the chiral theories presented in \cite{Lohitsiri:2019wpq}. These are certain $SU(N)\times Sp(r)$ gauge theories that exhibit a similar behaviour to the theories studied in \ref{sec:Georgi}. The theory can be studied in two limits. In the one flavor case, the two limits give the same massless spectrum and the same symmetry breaking pattern, which is consistent with having no phase transition as one varies the parameters. In the multi-flavour version of the theory, the two limits give different spectrum with different symmetries in the IR which means that there must be a phase transition. Exactly as in section \ref{sec:Georgi}, we will give a simple gauge invariant description with some condensates and massless fermions that hold for every number of flavours. 
The one flavor theory contains the left handed fermions in the following $SU(N)\times Sp(r)$ representations:
\eq{\psi\ \text{in the}\ (\Box,\Box)\ , \ \xi\ \text{in the}\ (1,\Box)\ ,\  2r\ \text{copies of}\ \eta\ \text{in the}\ (\overline{\Box},1)\ .} We demand that $N$ is odd to guarantee cancellation of the $\Z_2$ gauge anomaly of $Sp(r)$. Notice that for the minimal choice of $N=3,\ r=1$ the theory is an $SU(3)\times SU(2)$ sector of the standard model. The theory enjoys a global symmetry of $SU(2r)_\eta\times U(1)$ where under the $U(1)$, \eq{q_\psi=1\ ,\ q_\xi=-N\ ,\ q_\eta=-1\ .}
For the multiflavour version, we simply take $K$ copies of the fermion content. The global symmetry in this case is
\eq{SU(K)_\psi\times SU(K)_\xi\times SU(2rK)_\eta\times U(1)\ .}
Assuming that the theory confines at low energies, we give the following proposal for the low energy effective theory. The composite gauge invariant fermions $\Psi=\eta\psi\xi$ are massless in the IR. There are $2rK^3$ such fermions with $U(1)$ charge $q_\Psi=-N$. $U(1)^3$ and $U(1)$ gravitational anomalies imply that we should have only $2rK$ such fermions so some symmetry must be broken (assuming the $U(1)$ is not broken which is true in the two limits as shown in \cite{Lohitsiri:2019wpq}). Consider the two gauge invariant scalar operators $\mathcal{B}_N=(\psi\eta)^2$ and $\mathcal{B}_r=\psi^N\xi$. The condensate of $\mathcal{B}_N$ breaks the global symmetry to
\eql{symstrong}{SU(K)_{\psi\eta}\times SU(K)_\xi\times Sp(r)_\eta\times U(1)\ .}
The fermions $\Psi$ are then in the $(1,\Box,\Box)_{-N}$ under \eqref{symstrong}. This is exactly the phase obtained in \cite{Lohitsiri:2019wpq} in the $\Lambda_N\gg\Lambda_r$ limit. The condensate of $\mathcal{B}_r$ breaks the global symmetry to
\eql{symweak}{SO(K)_{\psi\xi}\times SU(2rK)_\eta\times U(1)\ .}
The fermions $\Psi$ are then in the $(1,\Box)_{-N}$ under \eqref{symweak}. This is exactly the phase obtained in \cite{Lohitsiri:2019wpq} in the $\Lambda_r\gg\Lambda_N$ limit.
Exactly as in \ref{sec:Georgi}, we propose that for intermediate $\Lambda_N/\Lambda_r$ the two operators condense. Together they break the symmetry down to
\eql{symboth}{SO(K)_{diag}\times Sp(r)\times U(1)\ ,}
with $\Psi$ in the $(\Box,\Box)_{-N}$ under \eqref{symboth}.

\section*{Acknowledgments}
We would like to thank Adi Armoni, Pietro Benetti Genolini, Philip Boyle Smith, Joe Davighi, Zohar Komargodski, Nakarin Lohitsiri, Kaan Onder, and Shimon Yankielowicz for many useful discussions. We would also like to especially thank David Tong for many discussions, insights and going over the draft. The author is supported by David Tong's Simons Investigator Award. This work has been partially supported by STFC consolidated grant ST/T000694/1

\appendix

\section{QCD and SPT phases}
The bifundamental theory that was described in \ref{sec:YM} and studied in \cite{Karasik:2019bxn} has nice applications to SPT classification of QCD. To explain the SPT classification of QCD, we will start from 4d $SU(N_c)$ pure YM theory. The theory enjoys a $\mathbb{Z}_{N_c}$ 1-form symmetry \cite{Gaiotto:2014kfa}. The 1-form symmetry can be coupled to a background gauge field $B$. The partition function then satisfies\cite{Gaiotto:2017yup}
\eql{phase}{\frac{Z[\te+2\pi k]}{Z[\te]}=exp\left(\frac{\pi i k(N_c-1)}{N_c}\int_{\mathcal{M}_4}B\cup B\right)\ ,}
where on spin manifolds, 
\eq{exp\left(\pi i \int_{\mathcal{M}_4}B\cup B\right)=1\ .}
The classification on non-spin manifold is different but we will restrict attention to spin manifolds for simplicity. The vacuum of YM for generic $\te$ is unique and trivially gapped, with first order phase transitions at $\te=\pi\mod 2\pi $. Vacuum states with different phase \eqref{phase} are in different SPT phases in the sense that they cannot be smoothly connected while remaining trivially gapped and without breaking the 1-form symmetry. States with the same phase, such as the vacuum of $\te=0$ and $\te=2\pi N_c$ can be connected smoothly without breaking the $\mathbb{Z}_{N_c}$ 1-form symmetry. This can be done for example by adding a massive adjoint fermion. This deformation doesn't break the 1-form symmetry, and the theory is still trivially gapped. Denoting the mass of the fermion by $M$, Shifting $\te$ by $2\pi$ is equivalent to rotating $M\to e^{2\pi i/N_c}M$. We immediately see that $\te=0$ and $\te=2\pi N_c$ are indistinguishable in the presence of an adjoint fermion. All the YM vacuum states in the same SPT phase collapse into one state in adjoint QCD. Indeed, the vacuum structure of light ($|M|\ll\Lambda$) adjoint QCD can be described by $N_c$ discrete states in which the chiral condensate is $\vev{\tilde{\psi}\psi}\sim e^{2\pi ik/N_c}$. For $|M|\ll\Lambda$ the $k$-vacuum has energy \eq{E_k\sim 1-cos\left(\frac{2\pi k}{N_c}+arg(M)\right)=1-cos\left(\frac{2\pi k+\te}{N_c}\right)\ .} This theory is strictly symmetric under $\te\to \te+2\pi N_c$. 
What happens if instead we add $N_f$ fundamental fermions? This deformation breaks the 1-form symmetry completely and one might think that the whole classification disappears. However, some of the classification survives thanks to an anomaly between the $2\pi$ periodicity of $\te$ and the faithful global symmetry $\frac{U(N_f)}{\mathbb{Z}_{N_c}}$ that was described in section \ref{sec:reviewQCD}. This anomaly exists only when $L\equiv gcd(N_c,N_f)>1$ which raises the question whether when $L=1$ we can connect the $\te=0$ and the $\te=2\pi$ vacua while remaining trivially gapped. 
Within QCD it looks like there is no way to do it, but it might be done by deforming the theory in the spirit of section \ref{sec:YM}. Consider weakly gauging the $SU(N_f)$ flavour symmetry and continuously increasing the $SU(N_f)$ gauge coupling. The theory we get is the massive deformation of the bifundamental theory mentioned in \ref{sec:YM} which was studied thoroughly in \cite{Karasik:2019bxn}. As explained, in the limit where $\Lambda_{c}\gg\Lambda_f$ which is our starting point, there are $N_f$ local minima classified by an integer $k$. This integer is related to the phase of the fermion bilinear condensate $U\sim e^{\frac{2\pi ik}{N_f}}$. The effective theory in each vacuum is $SU(N_f)$ YM with effective $\te$ angle $\te^{(k)}=\frac{2\pi N_c k}{N_f}$ where we took the uv $\te$ angles to be $0$. In addition, the mass of the fermion (which we take to be real and positive) induces a potential proportional to $E_{M,k}\sim -Mcos\left(\frac{2\pi k}{N_f}\right)$. Therefore, there is a unique groundstate which is the $k=0$ vacuum. changing $\te$ by $2\pi$ is equivalent to taking $k\to k-1$. Now we can continuously increase the value of $\Lambda_f$ until $\Lambda_f\gg\Lambda_c$. It has been argued in \cite{Karasik:2019bxn} that along this trajectory, the $\te=0$ theory remains trivially gapped. In the $\Lambda_f\gg\Lambda_c$ limit, the vacuum structure looks the same with the replacement of $N_c\leftrightarrow N_f$. There are $N_c$ local minima parametrized by an integer $k'$. How are the $N_f$ $k$-vacua mapped to the $N_c$ $k'$-vacua as we perform this deformation? Naively, one would say that $k=0$ is mapped to $k'=0$, $k=1$ to $k'=1$ and so on. However, $k,k'$ are defined mod $N_f,N_c$ which makes this mapping ambiguous. As a result, a certain $k$-vacuum is mapped to $k'=k\mod L$. Stated differently, all the $k$-vacua of QCD that differ by $L\mathbb{Z}$ are mixed in agreement with the $\Z_L$ SPT classification. 

\section{$\Z_N$ anomalies on junctions}
\label{sec:junctionanomalies}
As reviewed in the introduction, there are several types of anomalies associated with a broken $\Z_N$ symmetry. Some of them are realized on domain walls connecting two vacua. This can happen when the anomaly is of the form $\Z_N\times G^2$ where $G$ is an unbroken symmetry transformation and $G^2$ can describe a 3d anomaly. That's the situation for example in super Yang-Mills \cite{Dierigl:2014xta} and pure Yang-Mills at $\te=\pi$ \cite{Gaiotto:2017yup} where $G$ is a discrete one-form symmetry. It also happens for the $\te$ periodicity anomaly in QCD as reviewed in \ref{subsec:theta}, and for $N_f=1$ QCD \ref{subsec:cusp} where $G$ is a zero-form symmetry with a discrete quotient. 
The realization of $\Z_N$ anomalies on junctions is less familiar and we will review here the general way it works. It will be convenient to analyse a weakly interacting toy model which contains a complex scalar $\Phi=\rho e^{i\phi}$ and a Weyl fermion $\psi$ with the Yukawa interaction $\Phi\psi\psi+c.c$. At this stage, there is a $U(1)$ symmetry acting as \eql{u1}{U(1):\ \psi\to e^{i\al}\psi\ ,\ \phi\to\phi-2\al\ .}
This $U(1)$ is axial and has $U(1)\times gravity$ as well as $U(1)^3$ anomalies.
We can add a potential which is minimized by $\rho=v$. The $U(1)$ is spontaneously broken in the vacuum. The low energy theory contains only the $U(1)$ Goldstone boson $\phi$. How are the anomalies realized in this phase? This $U(1)$ sigma model has vortex-string configurations where $\phi$ winds around the string $\oint d\phi=2\pi$. At the core of the string, $\rho$ must go to zero. It was shown in \cite{Callan:1984sa} that there is a fermion zero mode in the effective 2d theory living on the string. This 2d fermion has pure gravity and $U(1)^2$ anomalies. Starting from $U(1)\times gravity$ and $U(1)^3$ in 4d, reducing to 2d on a vortex string gives the 2d anomalies $gravity$ and $U(1)^2$. It works exactly the same if we replace the $U(1)$ by a discrete subgroup. We can do it for example by adding a potential of the form $\Phi^N+\Phi^{*N}\sim cos(N\phi)$ \footnote{The fact that it is irrelevant for $N>4$ is irrelevant for our analysis}. This potential breaks explicitly $U(1)\to \Z_{2N}$ acting as
\eql{Z2N}{\Z_{2N}:\ \psi\to e^{\frac{\pi i}{N}}\psi\ ,\ \phi\to\phi-\frac{2\pi}{N}\ .} In addition, the vacuum breaks spontaneously $\Z_{2N}\to \Z_2$. This theory has $\Z_{2N}^3$ and $\Z_{2N}\times gravity$ anomalies. These anomalies cannot be realized on domain walls connecting two vacua. Naively we would expect to find pure gravitational anomaly and $\Z_{2N}^2$ anomaly on the domain wall to match the anomalies from the bulk. However, there are no such anomalies in 3d. Moreover, $\Z_{2N}$ is not even a symmetry of the domain wall theory. The $U(1)$ case we analysed before tells us exactly where to find these anomalies. We should take the $\Z_{2N}$ analogue of a vortex-string. Again we take $\phi$ to wind, but because the vacuum is only for discrete values $\phi_k=\frac{2\pi k}{N}$, the configuration we get is a junction of domain walls similar to figure \ref{pizza}. On the junction, the discrete broken symmetry is restored. The 2d theory is expected to carry the same anomalies as in the $U(1)$ case: $\Z_{2N}^2$ and $gravity$. We see that for a broken discrete symmetry, anomalies are realized on domain walls only in limited set of examples. In general they are realized on 2d junctions, similar to the case of broken $U(1)$.
The same mechanism described here was used in section \ref{subsec:Nf1baryons} in the construction of $N_f=1$ baryons, and can be applied for the theory studied in \ref{sec:AS} as was explained in \ref{subsec:mixing}.

\section{Anomalies of the multiflavour chiral theory}
\label{multianomalies}
In this appendix, we will analyse the anomalies of the theory studied in \ref{subsec:multiAS}. We will show that the proposed phase matches the anomalies, and in addition, that if the $U(1)$ is not broken, \eqref{breaking} is the minimal breaking pattern consistent with anomalies.

There are many anomalies that should be matched by the IR theory which are summarized in the following list:

\begin{itemize}
	\item $U(1)^3:\ -\frac{KN^3(N-4)(N-3)}{2}$
	\item $grav^2\times U(1):\ -\frac{KN(N-4)(N-3)}{2}$
	\item $SU(K)_\chi^3:\ \frac{N(N-1)}{2}$
	\item $SU(K(N-4))_\psi:\ N$
	\item $SU(K)_\chi^2\times U(1):\ \frac{N(N-1)(N-4)}{2}$
	\item $SU(K(N-4))_\psi^2\times U(1):\ N(2-N)$
\end{itemize}
It will be useful to define the subgroup $SU(K)_\psi\times SU(N-4)_\psi\subset SU(K(N-4))$.  We will also define $SU(K)_{diag}$ which is a diagonal combination of $SU(K)_{\chi}$ and $SU(K)_{\psi}$. Under $SU(K)_{diag}$ there are $\frac{N(N-1)}{2}$ fundamentals with charge $q=N-4$, and $N(N-4)$ antifundamentals with $q=2-N$. Under $SU(N-4)_\psi$ there are $KN$ fundamentals with charge $q=2-N$. The anomalies involving these groups are
\begin{itemize}
	\item $SU(K)_{diag}^3:\ -\frac{(N-4)(N-3)}{2}+6$
	\item $SU(K)_{diag}^2\times U(1):\ -\frac{N(N-4)(N-3)}{2}$
	\item $SU(N-4)_\psi^3:\ KN$
	\item $SU(N-4)_\psi^2\times U(1):\ KN(2-N)$
\end{itemize}

Assuming that $U(1)$ is not broken, the triangle $U(1)^3$ and gravitational anomalies must be matched in the IR by massless gauge invariant fermions. Consider the composite fermion $\chi\psi\psi$. It has charge $-N$ under the $U(1)$. To saturate the $U(1)^3$ and gravitational anomalies, we need $\frac{K(N-4)(N-3)}{2}$ such fermions. This number of fermions doesn't fit nicely into representations of $SU(K)_\chi\times SU(K(N-4))_\psi$. Therefore, some of the symmetries must be broken. We can try a phase in which $SU(K)_\chi$ is preserved and there are $\frac{(N-4)(N-3)}{2}$ fermions in the fundamental of $SU(K)_\chi$. This doesn't work since these fermions don't give the correct anomalies for $SU(K)_\chi^3$ and $SU(K)_\chi^2\times U(1)$. Therefore, $SU(K)_\chi$ must be broken. Instead we can assume that $SU(K)_{diag}$ is preserved and that we have $\frac{(N-4)(N-3)}{2}$ fermions in the fundamental of $SU(K)_{diag}$. Here the situation is a little bit better. The $SU(K)_{diag}^2\times U(1)$ anomaly is matched correctly, but we are missing 6 fermions for the matching of $SU(K)_{diag}^3$. For $K=2$, there is no $SU(K)_{diag}^3$ anomaly, only the $\Z_2$ valued Witten anomaly, but $6$ is even so we are fine. For $K>2$, $SU(K)_{diag}$ must be broken. The matching of $SU(K)_{diag}^2\times U(1)$ suggests that a breaking of $SU(K)_{diag}$ to some real subgroup (i.e. free from a triangle anomaly with itself) will do the job. A simple candidate is $SO(K)_{diag}$.  Now lets analyse the $SU(K(N-4))_\psi$ symmetry. The fermions can easily fit into $K$ fermions in the symmetric representation of $SU(N-4)_\psi$. Locking the $SU(K)_{\chi}$ and $SU(K)_{\eta}$ to give $SU(K)_{diag}$ breaks $SU(K(N-4))_\psi\to SU(N-4)_\psi$. These fermions match the anomalies for $SU(N-4)_\psi^3$ and $SU(N-4)_\psi^2\times U(1)$. To summarize, if we break 
\eq{SU(K)_\chi\times SU(K(N-4))_\psi\to SU(K)_{diag}\times SU(N-4)_\psi\ ,}
all the anomalies are matched except for $SU(K)_{diag}^3$ for $K>2$. If in addition we break it to some anomaly free subgroup such as $SO(K)_{diag}$, we are done. This is exactly the phase that we get by simply considering the same condensate $\vev{\mathcal{B}}\neq 0$ as in the $K=1$ case. 
\section{Bars-Yankielowicz theory}
\label{subsec:ASBY}
A famous generalization of the chiral theory introduced in \ref{subsec:basicAS} is known as the Bars-Yankielowicz (BY) theory \cite{Bars:1981se}. This is an $SU(N)$ gauge theory with an anti-symmetric\footnote{We discuss here only the anti-symmetric theory. The generalization to the symmetric BY theory is straight forward.} fermion $\chi$, $N-4+p$ anti-fundamental fermions $\psi$ and $p$ fundamental fermions $\xi$. Under general $U(1)$ transformations of the fermions
\eq{\chi\to e^{i\omega_\chi}\chi\ ,\ \psi\to e^{i\omega_\psi}\psi\ ,\ \xi\to e^{i\omega_\xi}\xi\ ,}
the $\te$ angle is shifted by
\eq{\te\to\te+(N-2)\omega_\chi+(N-4+p)\omega_\psi+p\omega_\xi\ .}
The global symmetry now is (ignoring discrete quotients)
\eq{G=SU(N-4+p)_\psi\times SU(p)_\xi\times U(1)_q\times U(1)_r\ ,}
where the charges of the fields under the two $U(1)$s are taken to be
\eq{&U(1)_q:\ q_\chi=N-4\ ,\ q_\psi=2-N\ ,\ q_\xi=N-2\\
	&U(1)_r:\ r_\chi=p\ ,\ r_\psi=0\ ,\ r_\xi=2-N\ .}
This theory has two main candidates for the IR phase \footnote{There are many more possibilities that involve larger symmetry breaking but we will focus only on the simplest two.}. The first is the phase in which no symmetry is broken and all the anomalies are matched by the composite massless fermions $\chi\psi\psi,\ \chi\psi\xi^\dagger,\ \chi\xi^\dagger\xi^\dagger$ \cite{Bars:1981se}. We will call this phase the BY phase. It was argued in \cite{Eichten:1985fs} that this phase cannot be realized in the large $N$, finite $p$ limit and proposed a different phase. The second phase involves the symmetry breaking pattern of \eq{SU(N-4+p)_\psi\times SU(p)_\xi\times U(1)_q\times U(1)_r\to SU(N-4)_\psi\times SU(p)_{\psi\xi}\times U(1)_q\times U(1)_{r'}\ .}
Here $SU(p)_{\psi\xi}$ is a diagonal combination of $SU(p)_\xi$ and $SU(p)_\psi\subset SU(N-4+p)_\psi$, and $U(1)_{r'}$ is a combination of $U(1)_r$ and the $SU(N-4+p)_\psi$ generator that commutes with $SU(N-4)_\psi\times SU(p)_{\psi\xi}$. The anomalies are matched in this phase partially by massless fermions and partially by Goldstone bosons. The condensate that breaks the symmetry is the fermion bilinear $\vev{\psi_i\xi^a}\sim\delta_{i}^a$ where $i=1,...,N-4+p$ and $a=1,...,p$ are the $SU(N-4+p),\ SU(p)$ indices respectively. The massless fermions are $\chi\psi\psi$ where now only the $N-4$ $\psi$ fermions charged under the unbroken $SU(N-4)_\psi$ participate in the construction of the massless fermions. We will call this phase the broken phase. As before, we can couple the theory to an additional $SU(N-1)$ gauge theory and a bifundamental Dirac fermion $\la_D$, and study the constraints coming from the $\Z_N$ axial symmetry. By repeating the arguments we conclude that the $\Z_N$ axial transformation that acts as\footnote{As explained in \eqref{basicaxial}, there is a $\Z_N$ transformation which is not embedded inside the other symmetries.} \eql{BYaxial}{\chi\to e^{-4\pi i/N^2}\chi\ ,\ \psi\to e^{2\pi i/N^2}\psi\ ,\ \xi\to e^{-2\pi i/N^2}\xi\ ,} carry some anomalies that must be matched somehow in the IR. One way to match the anomaly is by having a condensation of an operator charged under this $\Z_N$. We will discuss other possibilities later. In the BY phase we demand that this operator doesn't break any symmetry. There is a unique such operator which is $\mathcal{B}_{p}=\chi^{N-2}\psi^{N-4+p}\xi^p$. In the broken phase, we demand that this operator doesn't break any of the residual symmetries $SU(N-4)_\psi\times SU(p)_{\psi\xi}\times U(1)_q\times U(1)_{r'}$. This demand is weaker and allows a family of condensates $\mathcal{B}_m=\chi^{N-2}\psi^{N-4+m}\xi^m$. The simplest one is $\mathcal{B}_0=\chi^{N-2}\psi^{N-4}$. Interestingly these condensates shed a new light on the competition between the two phases. Applying weak interactions intuition to strongly interacting theories is not necessarily a wise thing to do. Yet, ideas based on most attractive channel and simplest condensate are commonly used in the literature. Naively, one would think that the BY phase is the simplest since it involves no symmetry breaking and therefore requires no condensates, while the broken phase involves the bilinear condensate $\psi\xi$. Other phases not discussed here involve different patterns of symmetry breaking and more complicated condensates. Our analysis flips the order of "naive attractiveness" since the BY phase requires a condensate made out of $2N-6+2p$ fermions. On the other hand, the broken phase requires in addition to the bilinear condensate $\psi\xi$, the condensate of $\mathcal{B}_0$ which is made out of only $2N-6$ fermions. So far the only possibility for matching the $\Z_N$ constraints we discussed is via a condensate. However for generic $\{N,p\}$ in the BY phase there is another option. The anomaly can be matched in principle via massless fermions. From \eqref{BYaxial} we see that the three massless fermions in the IR $\chi\psi\psi,\ \chi\xi^\dagger\xi^\dagger\ ,\ \chi^\dagger\psi^\dagger\xi$ are invariant under the $\Z_N$. We can attach $\mathcal{B}_p$ to one of the fermions. This will not change any of its quantum numbers but will change its $\Z_N$ transformation law. For example, replace $\chi\xi^\dagger\xi^\dagger$ with the fermion $\chi^{N-1}\psi^{N-4+p}\xi^{p-2}\simeq\mathcal{B}_p\chi^\dagger\xi^\dagger\xi^\dagger$  \footnote{$\simeq$ here means up to $(\xi^\dagger\xi)^2$ which is completely neutral.}. With this replacement the anomaly can be matched by the massless fermions without need of a condensate, but with the price of having a massless complicated composite fermion instead of the simple three-fermion operator. It is important to comment that this trick cannot work in the other examples mentioned in the paper. The reason is that the $\Z_N$ can be mixed with $U(1)$ symmetries. In the BY phase there are three massless fermions, and only two $U(1)$ symmetries, so in general we cannot take them all to be $\Z_N$ invariant. In the broken phase and in the other examples discussed here, only with the condensate we have more objects in the IR than $U(1)$ symmetries. Assuming that the BY phase is realized for some $\{N,p\}$, it is a hard question to tell which of the two scenarios is the correct one. This point will be further discussed from a Higgs phase perspective. The broken phase of the BY theories can easily reached by adding a scalar field with the Yukawa interaction $\phi\psi\xi$. $\phi$ is gauge invariant and sits in the $(\overline{\Box},\overline{\Box})_{0,N-2}$ under $SU(N-4+p)_\psi\times SU(p)\times U(1)_q\times U(1)_r$. Its vev breaks the symmetry as anticipated in the broken phase. Except for the NLSM caused by the symmetry breaking, there is still the $SU(N)$ gauge group coupled to $\chi$ and $N-4$ $\psi$s. This sector is nothing but the basic anti-symmetric theory. After confinement of the $SU(N)$, we get from this sector the massless fermions $\chi\psi\psi$ and a condensate of $\mathcal{B}_0=\chi^{N-2}\psi^{N-4}$.  
To get the BY phase using a weakly interacting trajectory, one has to work a little bit harder. At least for $p\leq 3$, there is a straight forward Higgsing that leads to the BY phase. This is done by simply adding the Higgs field $\phi$ in the $(\overline{\Box},\overline{\Box},1)_{2,-p}$ under $[SU(N)]_{gauge}\times SU(N-4+p)_\psi\times SU(p)_\xi\times U(1)_q\times U(1)_r$, with the Yukawa interaction $\phi\chi\psi+c.c$. Notice that if $p\geq 4$, the vev of $\phi$ breaks some of the global symmetries. For $p\leq 3$ it works almost exactly as in section \ref{subsec:mixing}. The vev of $\phi$ Higgses the gauge group down to $SU(N)\to SU(4-p)$, and locks the global symmetries with some gauge transformations. The massless fermions that remain under $[SU(4-p)]_{gauge}\times SU(N-4+p)\times SU(p)\times U(1)_q\times U(1)_r$ are
\begin{itemize}
	\item One fermion in the $(1,\sym,1)_{-N,p}$. It comes from $\psi$. It has the same quantum numbers as $\chi\psi\psi$. It is invariant under the $\Z_N$.
	\item One fermion in the $(1,\overline{\Box},\Box)_{N,2-N-p}$. It comes from $\xi$ and has the same quantum numbers as $\chi^\dagger\psi^\dagger\xi$. It is invariant under the $\Z_N$.
	\item One fermion $\chi'$ in the $\left(\anti,1,1\right)_{-pN/(4-p),p(p-4+2N)/(4-p)}$. It comes from $\chi$. It transforms under the $\Z_N$ as $\chi'\to e^{-\frac{4\pi i}{N(4-p)}}\chi'$.
	\item One fermion $\xi'$ in the $(\Box,1,\Box)_{N(2-p)/(4-p),4N/(4-p)-2N-p+2}$. It also comes from $\xi$. It transforms under the $\Z_N$ as $\xi'\to e^{-\frac{2\pi i}{N(4-p)}}\xi'$.
\end{itemize}
As we continue flowing down, the residual $SU(4-p)$ gauge theory will become strongly coupled. The gauge sector $SU(4-p)+\chi'+\xi'$ in the IR should turn into the missing ingredients for the BY phase.  
The main missing ingredient is a massless fermion in the $\left(1,\overline{\anti}\right)_{-N,p-4+2N}$ under $SU(N-4+p)\times SU(p)\times U(1)_q\times U(1)_r$. This fermion can be $\chi\xi^\dagger\xi^\dagger$ which is invariant under the $\Z_N$. In this case we need the $\mathcal{B}$ condensate. Alternatively, the fermion can be $\mathcal{B}_p\chi\xi^\dagger\xi^\dagger$ which is charged under the $\Z_N$. In this case the condensate is not needed.
At this point we should separate into cases:
\begin{enumerate}
	\item $p=1$: $\chi'$ and $\xi'$ are in the $(\overline{\Box},1)_{-N/3,\ 2N/3-1}$ and $(\Box,1)_{N/3,\ 1-2N/3}$ under $[SU(3)]_{gauge}\times SU(N-1)\times U(1)_q\times U(1)_r$ ($SU(p)=SU(1)$ is trivial of course). This is simply $N_f=1$ QCD. It is gapped with the $\xi'\chi'$ condensate which is neutral under all the symmetries but transforms under the $\Z_N$ as $\xi'\chi'\to e^{-\frac{2\pi i}{N}}\xi'\chi'$. The $\chi\xi^\dagger\xi^\dagger$ fermion doesn't exist in this case because there is no antisymmetric representation for $SU(p)$.
	\item $p=2$: $\chi'$ and $\xi'$ are in the $(1,1,1)_{-N,2N-2}$ and $(\Box,1,\Box)_{0,0}$ under $[SU(2)]_{gauge}\times SU(N-2)\times SU(2)\times U(1)_q\times U(1)_r$. In this case $\xi'\xi'$ condenses. It is neutral under all the symmetries but transforms under the $\Z_N$ as $\xi'\xi'\to e^{-\frac{2\pi i}{N}}\xi'\xi'$. We are left with the massless fermion $\chi'$ which has the same quantum numbers as $\chi\xi^\dagger\xi^\dagger$.
	\item $p=3$: In this case there is no gauge group. Similarly, $\chi'$ doesn't exist. We are left just with the massless fermion $\xi'$ in the $(1,\Box)_{-N,2N-1}$ which has the same quantum numbers as $\chi\xi^\dagger\xi^\dagger$ (recall that for $SU(3)$, $\Box\equiv \overline{\anti}$). However, unlike $\chi\xi^\dagger\xi^\dagger$ which is invariant under $\Z_N$, $\xi'$ transforms as $\xi'\to e^{-\frac{2\pi i}{N}}\xi'$.
\end{enumerate} 
In all three cases we have the correct massless fermion. For $p=1,2$ there is a condensate charged under $\Z_N$ in addition to the fermion. For $p=3$ there is no condensate, but the fermion is charged under $\Z_N$ so this phase is also consistent. One may wonder about the lack of universality and why $p=3$ is different than $p=1,2$. A possible reason is that the Higgs phase must give a consistent phase for every $N$ and in particular make sense also for pathological cases. For $p=1$, $N=3$ is a pathological case. This case is actually $N_f=1$ QCD in disguise. In this case there are no anomalies and no fermions in the IR. The $\Z_N$ constraints must be matched via a condensate. For $p=2$, $N=2$ is a pathological case. There is only one massless fermion in the IR. Using $U(1)$ transformations it can always be taken to be $\Z_N$ invariant. Also here there must be a condensate. For $p=3$ there are no pathological cases and the three IR fermions exist for every $N$. In this case, a phase without a condensate is possible. Because the Higgs phase must cover for all the possible cases, for $p=1,2$ a condensate must appear. For $p=3$ this demand is relaxed and indeed the Higgs phase doesn't exhibit a condensate. These subtleties provide another evidence supporting the importance of the $\Z_N$ constraints emphasized in this work.

	\bibliography{anomaly}

\providecommand{\href}[2]{#2}\begingroup\raggedright\begin{thebibliography}{10}

\bibitem{tHooft:1979rat}
G.~'t~Hooft, ``{Naturalness, chiral symmetry, and spontaneous chiral symmetry
  breaking},'' {\em NATO Sci. Ser. B} {\bf 59} (1980) 135--157.

\bibitem{Adler:1969gk}
S.~L. Adler, ``{Axial vector vertex in spinor electrodynamics},'' {\em Phys.
  Rev.} {\bf 177} (1969) 2426--2438.

\bibitem{Bell:1969ts}
J.~S. Bell and R.~Jackiw, ``{A PCAC puzzle: $\pi^0 \to \gamma \gamma$ in the
  $\sigma$ model},'' {\em Nuovo Cim. A} {\bf 60} (1969) 47--61.

\bibitem{Callan:1984sa}
C.~G. Callan, Jr. and J.~A. Harvey, ``{Anomalies and Fermion Zero Modes on
  Strings and Domain Walls},'' {\em Nucl. Phys. B} {\bf 250} (1985) 427--436.

\bibitem{Eichten:1985fs}
E.~Eichten, R.~D. Peccei, J.~Preskill, and D.~Zeppenfeld, ``{Chiral Gauge
  Theories in the 1/n Expansion},'' {\em Nucl. Phys. B} {\bf 268} (1986)
  161--178.

\bibitem{Karasik:2019bxn}
A.~Karasik and Z.~Komargodski, ``{The Bi-Fundamental Gauge Theory in 3+1
  Dimensions: The Vacuum Structure and a Cascade},'' {\em JHEP} {\bf 05} (2019)
  144, \href{http://xxx.lanl.gov/abs/1904.09551}{{\tt 1904.09551}}.

\bibitem{Tanizaki:2018wtg}
Y.~Tanizaki, ``{Anomaly constraint on massless QCD and the role of Skyrmions in
  chiral symmetry breaking},'' {\em JHEP} {\bf 08} (2018) 171,
  \href{http://xxx.lanl.gov/abs/1807.07666}{{\tt 1807.07666}}.

\bibitem{Wess:1971yu}
J.~Wess and B.~Zumino, ``{Consequences of anomalous Ward identities},'' {\em
  Phys. Lett. B} {\bf 37} (1971) 95--97.

\bibitem{Witten:1983tw}
E.~Witten, ``{Global Aspects of Current Algebra},'' {\em Nucl. Phys.} {\bf
  B223} (1983) 422--432.

\bibitem{Cordova:2019uob}
C.~C\'ordova, D.~S. Freed, H.~T. Lam, and N.~Seiberg, ``{Anomalies in the Space
  of Coupling Constants and Their Dynamical Applications II},'' {\em SciPost
  Phys.} {\bf 8} (2020), no.~1 002,
  \href{http://xxx.lanl.gov/abs/1905.13361}{{\tt 1905.13361}}.

\bibitem{Gaiotto:2017tne}
D.~Gaiotto, Z.~Komargodski, and N.~Seiberg, ``{Time-reversal breaking in
  QCD$_{4}$, walls, and dualities in 2 + 1 dimensions},'' {\em JHEP} {\bf 01}
  (2018) 110, \href{http://xxx.lanl.gov/abs/1708.06806}{{\tt 1708.06806}}.

\bibitem{Veneziano:1979ec}
G.~Veneziano, ``{U(1) Without Instantons},'' {\em Nucl. Phys.} {\bf B159}
  (1979) 213--224.

\bibitem{Witten:1980sp}
E.~Witten, ``{Large N Chiral Dynamics},'' {\em Annals Phys.} {\bf 128} (1980)
  363.

\bibitem{DiVecchia:1980yfw}
P.~Di~Vecchia and G.~Veneziano, ``{Chiral Dynamics in the Large n Limit},''
  {\em Nucl. Phys.} {\bf B171} (1980) 253--272.

\bibitem{Komargodski:2018odf}
Z.~Komargodski, ``{Baryons as Quantum Hall Droplets},''
  \href{http://xxx.lanl.gov/abs/1812.09253}{{\tt 1812.09253}}.

\bibitem{Karasik:2020pwu}
A.~Karasik, ``{Skyrmions, Quantum Hall Droplets, and one current to rule them
  all},'' \href{http://xxx.lanl.gov/abs/2003.07893}{{\tt 2003.07893}}.

\bibitem{Karasik:2020zyo}
A.~Karasik, ``{Vector dominance, one flavored baryons, and QCD domain walls
  from the ''hidden'' Wess-Zumino term},'' {\em SciPost Phys.} {\bf 10} (2021)
  138, \href{http://xxx.lanl.gov/abs/2010.10544}{{\tt 2010.10544}}.

\bibitem{Kapustin:2014zva}
A.~Kapustin and R.~Thorngren, ``{Anomalies of discrete symmetries in various
  dimensions and group cohomology},''
  \href{http://xxx.lanl.gov/abs/1404.3230}{{\tt 1404.3230}}.

\bibitem{Gaiotto:2013gwa}
D.~Gaiotto, ``{Kazama-Suzuki models and BPS domain wall junctions in N=1 SU(n)
  Super Yang-Mills},'' \href{http://xxx.lanl.gov/abs/1306.5661}{{\tt
  1306.5661}}.

\bibitem{Dimopoulos:1980hn}
S.~Dimopoulos, S.~Raby, and L.~Susskind, ``{Light Composite Fermions},'' {\em
  Nucl. Phys. B} {\bf 173} (1980) 208--228.

\bibitem{Bolognesi:2020mpe}
S.~Bolognesi, K.~Konishi, and A.~Luzio, ``{Dynamics from symmetries in chiral
  $SU(N)$ gauge theories},'' {\em JHEP} {\bf 09} (2020) 001,
  \href{http://xxx.lanl.gov/abs/2004.06639}{{\tt 2004.06639}}.

\bibitem{Bolognesi:2021yni}
S.~Bolognesi, K.~Konishi, and A.~Luzio, ``{Probing the dynamics of chiral
  $SU(N)$ gauge theories via generalized anomalies},'' {\em Phys. Rev. D} {\bf
  103} (2021), no.~9 094016, \href{http://xxx.lanl.gov/abs/2101.02601}{{\tt
  2101.02601}}.

\bibitem{Smith:2021vbf}
P.~B. Smith, A.~Karasik, N.~Lohitsiri, and D.~Tong, ``{On Discrete Anomalies in
  Chiral Gauge Theories},'' \href{http://xxx.lanl.gov/abs/2106.06402}{{\tt
  2106.06402}}.

\bibitem{Anber:2021iip}
M.~M. Anber, S.~Hong, and M.~Son, ``{New Anomalies, TQFTs, and Confinement in
  Bosonic Chiral Gauge Theories},''
  \href{http://xxx.lanl.gov/abs/2109.03245}{{\tt 2109.03245}}.

\bibitem{KOT}
A.~Karasik, K.~Onder, and D.~Tong, ``{In progress},''.

\bibitem{Bolognesi:2021hmg}
S.~Bolognesi, K.~Konishi, and A.~Luzio, ``{Strong anomaly and phases of chiral
  gauge theories},'' {\em JHEP} {\bf 08} (2021) 028,
  \href{http://xxx.lanl.gov/abs/2105.03921}{{\tt 2105.03921}}.

\bibitem{Georgi:1985hf}
H.~Georgi, ``{A Tool Kit for Builders of Composite Models},'' {\em Nucl. Phys.
  B} {\bf 266} (1986) 274--284.

\bibitem{Lohitsiri:2019wpq}
N.~Lohitsiri and D.~Tong, ``{If the Weak Were Strong and the Strong Were
  Weak},'' {\em SciPost Phys.} {\bf 7} (2019), no.~5 059,
  \href{http://xxx.lanl.gov/abs/1907.08221}{{\tt 1907.08221}}.

\bibitem{Gaiotto:2014kfa}
D.~Gaiotto, A.~Kapustin, N.~Seiberg, and B.~Willett, ``{Generalized Global
  Symmetries},'' {\em JHEP} {\bf 02} (2015) 172,
  \href{http://xxx.lanl.gov/abs/1412.5148}{{\tt 1412.5148}}.

\bibitem{Gaiotto:2017yup}
D.~Gaiotto, A.~Kapustin, Z.~Komargodski, and N.~Seiberg, ``{Theta, Time
  Reversal, and Temperature},'' {\em JHEP} {\bf 05} (2017) 091,
  \href{http://xxx.lanl.gov/abs/1703.00501}{{\tt 1703.00501}}.

\bibitem{Dierigl:2014xta}
M.~Dierigl and A.~Pritzel, ``{Topological Model for Domain Walls in
  (Super-)Yang-Mills Theories},'' {\em Phys. Rev. D} {\bf 90} (2014), no.~10
  105008, \href{http://xxx.lanl.gov/abs/1405.4291}{{\tt 1405.4291}}.

\bibitem{Bars:1981se}
I.~Bars and S.~Yankielowicz, ``{Composite Quarks and Leptons as Solutions of
  Anomaly Constraints},'' {\em Phys. Lett. B} {\bf 101} (1981) 159--165.

\end{thebibliography}\endgroup

\end{document}